\documentclass[a4paper,fleqn,usenatbib]{mnras}

\usepackage{amsmath}	
\usepackage{newtxtext,newtxmath}
\usepackage[T1]{fontenc}
\usepackage{graphicx}
\usepackage{subcaption}	

\DeclareRobustCommand{\VAN}[3]{#2}
\let\VANthebibliography\thebibliography
\def\thebibliography{\DeclareRobustCommand{\VAN}[3]{##3}\VANthebibliography}

\title[Magnetic Field Evolution]{Magnetic Field Evolution for Keplerian and Sub-Keplerian Flows around Non Rotating Black Hole }

\author[M. Mousapour et al.]{
M. Mousapour Gharghabi,$^{1}$\thanks{E-mail: m.mousapour@khayyam.ac.ir}
J. Ghanbari,$^{1,2}$\thanks{E-mail: j.ghanbari@khayyam.ac.ir}
M. Moeen Moghaddas$^{3}$\thanks{E-mail: Dr.moeen@kub.ac.ir}
\\
$^{1}$Department of Sciences, Khayyam University of Mashhad \\
$^{2}$Faculty of Basic Sciences, Ferdowsi University of Mashhad\\
$^{3}$Department of Sciences, Kosar University of Bojnord
}

\date{Accepted XXX. Received YYY; in original form ZZZ}

\pubyear{2024}

\begin{document}
\label{firstpage}
\pagerange{\pageref{firstpage}--\pageref{lastpage}}
\maketitle

\begin{abstract}
The exact two-dimensional non-stationary solution for the evolution of the magnetic field during accretion with nearly spherical symmetry, using Newton's solution and considering both Keplerian and sub-Keplerian flows around the black hole (BH), has been derived using the Schwarzschild metric. In this paper, we also discuss the possible origins of large-scale magnetic field production around black holes (BHs). For example, the origin of this strong large-scale magnetic field could be the interstellar medium or a companion star outside the accretion disk, which is drawn in by the accretion plasma and grows over time. The findings show that the uniform magnetic field of a subsonic flow, which is weak at infinity, increases with time and evolves into a quasi-radial field. This is true for both types of flows. We have also observed that the growth of the radial component of the magnetic field is more pronounced in the sub-Keplerian flow than in the Keplerian flow. However, it is worth noting that, for both types of flows, the magnetic field does not reach its saturation value in the regions  $r> r_{c}$; instead, the process of strengthening and growing the magnetic field progresses to a point where the disk evolves from its initial condition to a state close to the magnetically arrested disk (MAD) state, forming a sub-MAD state.
\end{abstract}



\section{Introduction}
Astrophysical accretion discs exert a pervasive influence across a vast spectrum of scales, ranging from planetary formation processes to the evolutionary pathways of galaxies. Serving as the impetus behind some of the most energetic phenomena in the cosmos, these discs play a pivotal role in shaping celestial dynamics. For example, the luminosity of X-ray emissions emanating from accretion discs orbiting stellar-mass BHs and neutron stars is prominently distinguished among celestial sources \citep{2006ARA&A..44...49R}. Additionally, the immense power of active galactic nuclei (AGNs) originates from the accretion of matter onto supermassive BHs located at the centers of galaxies. This characterization not only designates them as the predominant sources in the universe but also underscores their crucial role in regulating the overall evolution of galaxies through the energy feedback mechanism they manifest \citep{2023ApJ...944..182D, 2017NatAs...1E.165H, 2017FrASS...4...42M}.

The presence of magnetic fields plays a crucial role in facilitating angular momentum transfer, jet formation, interactions between BHs and their surrounding accretion discs, as well as the initiation of various  instabilities within accretion flows \citep{2023ApJ...944..182D}. It is typically expected that all three components of magnetic fields exist within accretion discs, with these fields being augmented through differential rotation. Generally, two views are accepted among astrophysicists regarding the generation of magnetic fields: the Biermann Battery Mechanism and the magnetohydrodynamics (MHD) dynamo process, which we will explain in the continuation of the text. 

Biermann Battery Mechanism: This mechanism is known for its capability to generate seed magnetic fields from an initial field condition of zero. The operation of this mechanism requires specific temperature and density variations within the environment. Crucially, the gradients of these variations should be non-parallel, indicating that these two variables change independently of each other, with changes in one variable not directly influencing changes in the other. The non-parallel nature of temperature and density variations plays a pivotal role in the generating magnetic instabilities and amplifying magnetic fields in accretion environments. This characteristic leads to the automatic generation of primary magnetic fields, which, over time, evolve into more intricate and potent configurations \citep{1982PASP...94..627K}.

MHD dynamo process: This process explains the creation of magnetic fields by transferring energy and mass within a medium. These magnetic fields are first produced locally on a small scale, serving as the seeds of larger fields, and are gradually strengthened by more complex mechanisms and interactions on a larger scale \citep{1995ApJ...446..741B}. Another possibility involves the generation of large-scale magnetic fields originating external to the accretion disc. These fields may come from sources such as the interstellar medium or companion stars, which are subsequently drawn into the accretion plasma \citep{1974Ap&SS..28...45B, 2019Univ....5..146B}. Later, the applications of this strong magnetic field on a large scale were studied, and numerical results confirmed their role in the MAD regime, where the poloidal field on a large scale plays a central role \citep{2003PASJ...55L..69N, 2011MNRAS.418L..79T}. The strength of the magnetic field is crucial in the dynamics of accretion, and in the accretion process, the large-scale strong magnetic field plays a very important role in angular momentum transfer, radiation processes, and the formation of strong jets.
    
Astrophysicists believe that the magnetic field may be created by the action of a dynamo, potentially within the disc itself. Notably, dynamos serve as amplification mechanisms contingent upon the existence of seed magnetic fields. Conversely, an alternative perspective suggests that the initial weak magnetic field could be introduced from an external source. This external source could be the outer regions of the disc or a companion star in the case of X-ray binaries, or the environment around AGNs, with the field then being enhanced by flow freezing \citep{10.1093/mnras/stab3790, 2023ApJ...944..182D, 2021ApJ...909..158L, 2020ApJ...900...59M, 2022ApJ...935...22M, 1988PhRvD..37.2743T, 2021ApJ...911...85V}. 

Accretion discs have the capability to transport magnetic fields on both small and large scales. The small-scale magnetic field is locally generated through turbulence induced by the magneto-rotational instability. This type of magnetic field can efficiently transfer angular momentum outward across a significant region of the disc. In contrast to the small-scale field, the large-scale magnetic field is likely not generated during the accretion process itself. Instead, it may either be acquired from the surrounding environment, drawn in by the accretion flow, or inherited from the past evolutionary history of the accreting flow. Nevertheless, the origin and strength of large-scale magnetic fields during the process of BH accretion remain inadequately understood. However, there must be an upper limit to the amount of magnetic flux that a disc around a BH can carry. Unlike small-scale magnetic fields, the large-scale fields cannot be neglected due to the minimal plasma resistance. Additionally, it is worth noting that a central BH does not absorb these fields \citep{2003ApJ...592.1042I}. 

If a BH can successfully transport the magnetic field lines to smaller radii and maintain the flux, it eventually encounters a critical threshold where the magnetic pressure resists continued accretion \citep{1974Ap&SS..28...45B, 2019Univ....5..146B, 2003PASJ...55L..69N, 2011MNRAS.418L..79T}. When a sufficient quantity of poloidal magnetic flux is present, it leads to the formation of an extremely efficient mechanism called a MAD \citep{2003PASJ...55L..69N}. Recent validations through general relativistic magnetohydrodynamic (GRMHD) simulations confirm this prediction \citep{10.1093/mnras/stab3790, 2014MNRAS.440.2185D, 2003ApJ...592.1042I, 2011MNRAS.418L..79T}. In this model, the accreting gas is compelled to migrate toward the central region within a robust polar magnetic field, resulting in the accumulation of magnetic flux that disrupts the axisymmetric accretion flow at a relatively large radius. Within the perturbation radius, the flow undergoes fragmentation into bubbles, forcing the gas to be driven toward the BH through the process of magnetic reconnection \citep{2019Univ....5..146B, 2003PASJ...55L..69N}. Under ideal circumstances, the magnetic field in the vicinity of a BH initially lacks the requisite strength, allowing unimpeded gas accretion. As accretion continues, the magnetic field progressively accumulates, leading to an increase in magnetic pressure near the BH. Consequently, over time, the magnetic pressure required to disturb the flow is eventually generated, causing the gas velocity to significantly decrease compared to the free-fall velocity, ultimately leading to the formation of the MAD state. The MAD disc comprises two distinct components: An outer accretion disc that is roughly axisymmetric and an inner accretion disc that emerges within the magnetically dominated region. The transition between these two regions occurs at the magnetospheric radius, where the accumulated vertical magnetic field perturbs the outer disc \citep{2019Univ....5..146B}. The jet efficiency in such configurations surpasses 100\% for thick discs and reaches approximately 50\% for thin discs with an aspect ratio of  $\frac{H}{R} = 0.03$ \citep{2020MNRAS.494.3656L, 2012MNRAS.423.3083M, 2011MNRAS.418L..79T}.

The MAD model posits a relationship between the mass accretion rate, magnetic flux threading a BH, and jet power, aligning with observations of radio-loud AGNs \citep{2014Natur.515..376G, 2014Natur.510..126Z}. Recent polarization analyses of M87, utilizing data from the Event Horizon Telescope (EHT) observations, support the existence of a dynamically significant organized, poloidal magnetic flux near a BH, consistent with GRMHD models of MAD \citep{2021ApJ...910L..12E, Akiyama_2021, 2022ApJ...924..124Y}. In most numerical simulations of accreting discs, especially MAD, the accretion flow starts with a sufficiently strong large-scale poloidal flux, which subsequently migrates closer to the BH, accumulating via flow-freezing \citep{2012MNRAS.423.3083M, 2011MNRAS.418L..79T}. The process of flow accumulation near a BH depends on the relative efficiency between the inward advection by the accretion flow and the outward diffusion due to turbulent resistivity \citep{10.1093/mnras/267.2.235}. Simulations depicting large-scale accretion flows around the galactic center, fed by magnetized winds from Wolf-Rayet stars, illustrate the effective inward transport of magnetic fields towards the central region \citep{2020MNRAS.492.3272R, 2020ApJ...896L...6R}. Nevertheless, the source of the large-scale magnetic field and the origin of the magnetic flux in the vicinity of a BH are still not fully understood and  remain active areas of research.

In contrast to the abundance of simulation-based studies, there is a significant lack of theoretical research on the formation and evolution of magnetically arrested disks (MADs). Previous research has primarily relied on complex numerical simulations to investigate the dynamics of MADs, leaving a substantial gap in the theoretical understanding of these phenomena. The aim of this paper is to address this gap by focusing on the evolution of magnetic fields in Keplerian and sub-Keplerian flows around non-rotating BHs. This study helps provide a theoretical framework that can guide and validate future simulations. Emphasizing the theoretical approach, our work not only confirms and complements simulation-based research but also provides new insights into the magnetic processes shaping accretion flows. This approach underscores the importance of analytical methods in advancing our knowledge of MADs, leading to a better understanding of these magnetic structures and providing more accurate and predictive models in astrophysics.
This article is organized as follows. In Section \ref{sec:MHDFLOW}, we discuss stationary axisymmetric MHD flow in a Schwarzschild spacetime, including subsections on flow properties and basic formulae. In Section \ref{sec:Kepnonkep}, we address the matching equations for Keplerian and sub-Keplerian flows. In section \ref{sec:paraequ}, we discuss the parameterization of equations in the region where $r> r_{c}$. Finally, we examine the key points of our results in section \ref{sec:conclusion}.

\section{MHD FLOW IN A SCHWARZSCHILD SPACETIME}\label{sec:MHDFLOW}
\subsection{Flow Properties}\label{subsec:FlowProperties}
The problem is solved based on the following assumptions. 
As described by \citeauthor{2019Univ....5..146B}, we similarly divide the flow region into three distinct areas. The first region, surrounding the collapsing star $\left(r < r_{c}\right)$, exhibits a steady supersonic flow characterized by a relatively low thermal velocity. The second region maintains a constant flow but marks the transition from subsonic to supersonic flow. The third and farthest region $r> r_{c}$ is a subsonic area that remains almost spherically symmetric over time without affecting the overall flow pattern. Within the subsonic region, which is stationary relative to the magnetic field, stagnation zones form where matter does not accrete onto the star, and the magnetic field pressure prevails. Stationary conditions are determined by the hydrodynamic time $t_{s}  \sim   {r}/{v} \sim  r^{3/2}$,  at $v \sim r^{-1/2}$ in the supersonic region, and $t_{s} \sim  r^{3}$ at $v \sim r^{-2}$ in the subsonic region \citep{1974Ap&SS..28...45B, 2019Univ....5..146B}. In this article, we aim to investigate the growth of the magnetic field in the third region. Thus we consider an almost spherically symmetric subsonic flow that extends into regions far away from the non-rotating BH. We assume that at infinity, a weak magnetic field is frozen into the flow, which increases over time and begins to influence the flow \citep{1974Ap&SS..28...45B, 2019Univ....5..146B}, we use spherical coordinates, and since the BH is non-rotating, the Schwarzschild metric is a suitable candidate.
\begin{equation}
    ds^{2} = \left(1-\frac{r_{g} }{r}  \right)c^{2}dt^{2} - \left(1-\frac{r_{g} }{r} \right) ^{-1}dr^{2} - r^{2} \left(d\theta ^{2}+\sin^{2}\theta d\varphi ^{2} \right), 
    \label{eq:eq1}
\end{equation}
where $r_g=\frac{2GM}{c^2}$ is the gravitational radius in Schwarzschild metrics, $c$ is the speed of light, $G$ is the gravitational constant, and $M$ is the mass of the BH.

In recent studies, such as that by \citeauthor{2019Univ....5..146B}, the rotational velocity component of the inflowing material in the accretion flow has been neglected. While these studies provide valuable insights, they do not offer a detailed and comprehensive understanding of accretion processes and their astrophysical consequences. The rotational velocity component is a critical factor that affects energy distribution, momentum transfer, and magnetic field configuration in the accretion disk. The presence of rotational motion increases the kinetic energy of the infalling material, leading to higher thermal energies and enhanced radiation output due to increased viscous dissipation. Additionally, rotational motion helps conserve angular momentum within the disk, ensuring its stability and facilitating the efficient transport of material toward the BH. Moreover, the consideration of rotation intensifies the dynamo effect, producing stronger magnetic fields capable of launching powerful jets. This study aims to address this gap by including the rotational component in our calculations to provide a more comprehensive understanding of accretion phenomena, thereby elucidating the critical role of rotation in shaping the physical properties of accretion discs around BHs.
Therefore, considering that the rotational component of the velocity is a significant factor affecting the accretion environment, we include it in our analysis. Thus, the flow velocity is given by $u^{\mu} = \left(u^{0},u^{r},0,u^{\varphi}\right)$, where $( \mu =0,1,2,3)$.

Similar to the work of \citeauthor{2019Univ....5..146B}, we assume the magnetic field is poloidal, represented by $B^{\mu} = \left(B^{0},B^{r},B^{\theta },0\right)$, meaning it is homogeneous in distant regions and its effect on the flow is negligible. In our analysis, the four-dimensional electric field vector $E^{\mu }=F^{\nu \mu }u_{\nu }$ vanishes under the "frozen" condition \citep{1974Ap&SS..28...45B, 2019Univ....5..146B}. Consequently, similar to Kogan, we disregard the effects of the electric field in our study.

\subsection{Basic Formulae}\label{sect:BasicFormulae}
In this section, we derive and present the necessary formulas related to the magnetic flux induction equation.
In analyze the evolution of large scale poloidal flux we utilize the poloidal flux function, which corresponds to the $\varphi$-component of the magnetic vector potential. In GRMHD, we directly evolve the components of the Faraday tensor, $F_{lm}$ using the alternative form of the induction equation (for further details see \citep{2003ApJ...589..458D}.

The general form of the induction equation is given by $\triangledown _{\mu } F^{\mu \nu }=0$, where covariant derivatives can be replaced with simple coordinate derivatives, leading to the following expression:
\begin{equation}    
    \partial_{\gamma} F_{\alpha \beta}+\partial_{\alpha} F_{\beta \gamma}+\partial_{\beta } F_{\gamma \alpha } = 0.
    \label{eq:eq2}
\end{equation}
The space-space components of $F_{\delta \gamma }$ are identified with the constrained-transport (CT) magnetic fields via $B^{i}=\left[i,j,k\right]F_{jk}, \left(i,j,k=1,2,3\right) $:
\begin{equation}    
    B^{r}=F_{\varphi \theta },\quad B^{\theta }=F_{r\varphi },\quad B^{\varphi }=F_{\theta r}.
    \label{eq:eq3}
\end{equation}
This identification allows us torewrite the induction equation in the familiar form
\begin{equation}    
    \partial_{t} B^{i}-\partial_{j} (B^{i} u^{k}-B^{k} u^{i} )=0,\quad \left(i,j,k=1,2,3\right).
    \label{eq:eq4}
\end{equation}
To determine the time dependence of the magnetic field, we use the ideal MHD equation under the assumption that the flow moves radially in the gravitational field of a BH \citep{1974Ap&SS..28...45B}.
\begin{equation}    
    \frac{\partial}{\partial x^{\nu }}  \sqrt{-g} (B^{\mu } u^{\nu }-B^{\nu } u^{\mu } )=0,\quad \left(\mu , \nu =0,1,2,3\right),
    \label{eq:eq5}
\end{equation}
where the metric $g_{\mu\nu},\, \left|g_{\mu\nu}\right| = g$ are known, with $\sqrt{-g}=r^{2} \sin \theta$. 
The tensor $F_{\delta \gamma }$ is related to the magnetic induction in the flow frame, $B^{\mu}$, by the relation:
\begin{equation} 
    B^{\mu }=\frac{1}{2\sqrt{g}}  \varepsilon ^{\mu \nu \delta \gamma } u_{\nu } F_{\delta \gamma },\quad B^{\mu } u_{\mu }=0,\quad \left(\mu , \nu , \delta , \gamma =0,1,2,3\right) ,
    \label{eq:eq6} 
\end{equation}
where $B^{\mu}$ is the four-dimensional magnetic field vector.
The time-dependent magnetic field components and initial field configuration are given by:

\begin{equation} 
    \frac{d}{dt} (\sqrt{-g}  u_{0}^{-1} B^{r} (u_{r} u^{r}+u_{0} u^{0} ))=0, \quad \frac{d}{dt} (\sqrt{-g}  u_{r} B^{\theta } )=0.
    \label{eq:eq7}
\end{equation}
Also we have
\begin{equation} 
    \frac{d}{dt} = \frac{\partial}{\partial t}+c\frac{u^{r}}{u^0} \frac{\partial}{\partial r}.
    \label{eq:eq8}
\end{equation}

Equation~(\ref{eq:eq7}) is related to the conservation of magnetic flux along the radial and tangential directions. By integrating the set of relations~(\ref{eq:eq7}) and~(\ref{eq:eq8}), we obtain the magnetic flux relations as follows:

\begin{equation} 
    \sqrt{-g}  u_{0}^{-1} B^{r} (u_{r} u^{r}+u_{0} u^{0} )=c_{2} r_{g}^{2},\quad \sqrt{-g}  u_{r} B^{\theta }=c_{3} r_{g}^{2},
    \label{eq:eq9}
\end{equation}
\begin{equation} 
    ct-\int{\frac{dr}{\frac{u^{r}}{u^{0}}}}=c_{1} r_{g}.
    \label{eq:eq10}
\end{equation}
In relation~(\ref{eq:eq10}), the time $t$ explicitly enters the integrals, so the magnetic field at a given spatial point depends on time in a hydrodynamically stationary flow.
In equations~(\ref{eq:eq9}) and~(\ref{eq:eq10}), $c_{1}r_{g}$, $c_{2}r_{g}^2$ and $c_{3}r_{g}^2$ are constants of integration. Their calculation is discussed in Appendix A.

\section{KEPLERIAN AND NON-KEPLERIAN FLOW }\label{sec:Kepnonkep}

Static accretion in the Schwarzschild metric, without a magnetic field, has already been investigated by \citet{1972Ap&SS..15..153M}. In the hypersonic region $r < r_{c}$, where $r_{c}$ is the sound velocity radius, the velocity of matter can be approximated by free fall \citep{2019Univ....5..146B}. We derive the behavior of the magnetic field components for Keplerian and non-Keplerian flows in the distant regions from the BH, $r> r_{c}$, where the flow motion is rotational. It is necessary to rewrite the relations in two cases.

\subsection{Keplerian flow equations  }\label{sec:Kepequ}

We consider the flow to be Keplerian and obtain the four-dimensional velocity components for this flow from the integrals of motion, related to the conservation of energy $E$ and zero angular momentum $L = 0$, as well as the angular velocity relation for the Keplerian flow: 
\begin{equation}
\Omega = \frac{u^{\varphi }}{u^{t}} = \sqrt{\frac{GM}{r^{3}}} = \frac{c}{\sqrt{2}r_{g} }  x^{3/2}, 
    \label{eq:eq11}
\end{equation}
where $x=\frac{r_{g}}{r}$ and $r_{g}=\frac{2GM}{c^{2}}$.

Using the relations from \citet{1974Ap&SS..28...45B}, we have:
\begin{equation} 
    E=mc^{2} g_{00} u^{0}=mc^{2} \left(1-\frac{r_{g}}{r} \right) u^{0},\quad \sqrt{-g} =r^{2}\sin\theta.
    \label{eq:eq12}
\end{equation}
As a result, using the above relations for matter at rest at infinity, i.e., $E = mc^{2}$, we derive the velocity components for the Keplerian flow from Eq. (12), using $u^{\mu } u_{\mu }=1$ as follows: 
\begin{equation}
\begin{split}
    &u_{0}=1,\quad u^{0}=\left(1-x\right) ^{-1}, \\
    &u^{r}=-\left(x-\frac{x\sin ^{2}\theta }{1-x } \right) ^{1/2} ,\quad u_{r} = \frac{1}{1-x}  \left(x-\frac{xsin^{2}\theta }{1-x } \right)  ^{1/2}, \\
    & u^{\varphi }=\frac{x^{3/2}}{r_{g}\left(1-x\right) } ,\quad u_{\varphi }=-\frac{sin^{2}\theta}{1-x}\frac{r_{g}}{\sqrt{x} }.
\end{split}
\label{eq:eq13}
\end{equation}
Following \citeauthor{2019Univ....5..146B}'s approach, we also assume the initial field to be a uniform magnetic field along the $z$ axis. In the Schwarzschild coordinate system, the physical components of the magnetic field are given by:
\begin{equation}
\begin{split}
    &_r B=\sqrt{-g_{rr}  } B^{r}=\left(1-\frac{{r_g}}{r} \right)^{{-1}/{2}} B^{r}, \\
    &_\theta B =\sqrt{-g_{\theta \theta }  } B^{\theta }=rB^{\theta }.
\end{split}
\label{eq:eq14}
\end{equation}
The primary field components are:
\begin{equation}
\begin{split}
    &B^{r} \left(t=0\right) =B_{0}\cos\theta \sqrt{1-x_{0}}, \\
    &rB^{\theta } \left(t=0\right) =-B_{0}sin\theta \sqrt{1-x_{0}} .
\end{split}
\label{eq:eq15}
\end{equation}
By using the velocity components and relation~(\ref{eq:eq7}) of the magnetic flux, we extract the parametric components of the magnetic field for the Keplerian flow:
\begin{equation}
\begin{split}
    B^{\theta }= &-\frac{B_{0}}{r_{g}} \sin\theta \frac{x^{2}}{x_{0}} \sqrt{1-x_{0}} \\
    &           \left\{\frac{1-x}{1-x_{0}} \frac{x_{0}\left(1-x_{0}\right)-x_{0}\sin^{2}\theta  }{x\left(1-x\right)-x\sin^{2}\theta } \right\}^{1/2}, \\
    B^{r }=\ &B_{0} \cos \theta \frac{x^{2}}{x_{0}^{2}} \sqrt{1-x_{0}} \\ 
    &       \left\{\left(\frac{1-x}{1-x_{0}}\right)^{2}  \frac{\left(1-x_{0}\right) -x_{0}\left(1-x_{0}\right)+x_{0}\sin^{2}\theta  }{\left(1-x\right)- x\left(1-x\right)+x\sin^{2}\theta }\right\}.
\end{split}
\label{eq:eq16}
\end{equation}

\begin{figure*}
    \centering
    \includegraphics[width=0.9\textwidth]{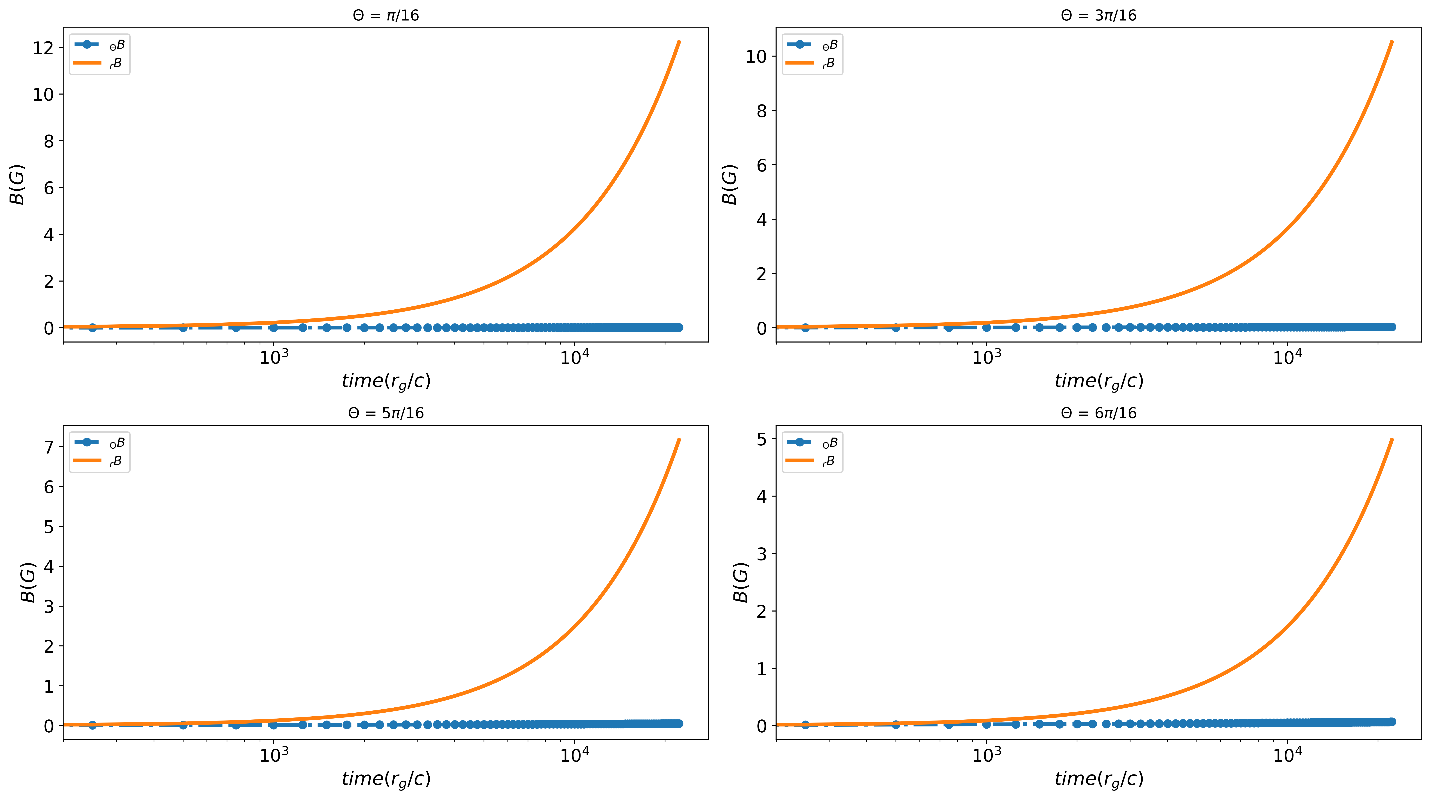}
    \caption{ Comparison of the time evolution of the physical components of the magnetic field in Keplerian flow at different polar angles.}
    \label{fig:fig1}
\end{figure*}

We plotted the temporal evolution of the physical components of the magnetic field in a Keplerian flow in Figure \ref{fig:fig1}. Each graph is plotted per angle, showing how the physical components of the magnetic field change with time for different angles. 
Additionally, to align with the time scales of magnetic field evolution investigated in previous studies, including the article by \citet{10.1093/mnras/stw1643}, the horizontal axis of our plots is considered up to the order of $10^4$. This time scale has been chosen to match simulation models and previous results, effectively demonstrating the dynamic changes of the magnetic field in accretion flows.

Figure \ref{fig:fig1} shows that as the angle $\theta $ increases, there is a continuous pattern in the growth of $_r B $ and the relative stability of $_\theta B $. The magnitude of $_r B $ grows faster for larger angles, while $_\theta B $  remains relatively stable and does not show significant changes.

\begin{figure*}
    \centering
    \begin{subfigure}{0.45\textwidth}
        \includegraphics[width=\textwidth]{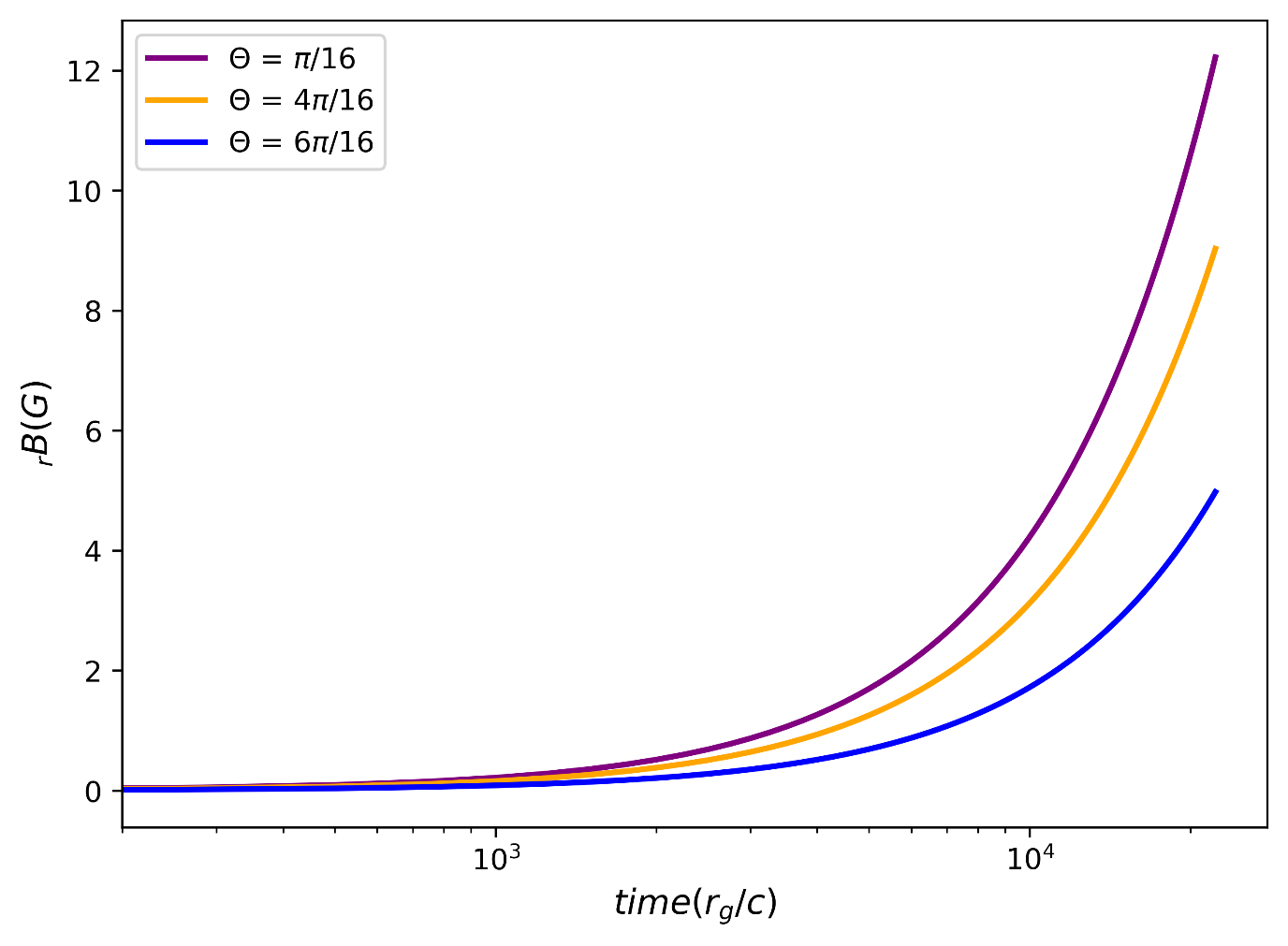}
        \caption{}
        \label{fig:fig2a}
    \end{subfigure}
    \hspace{-0.01\textwidth} 
    \begin{subfigure}{0.45\textwidth}
        \includegraphics[width=\textwidth]{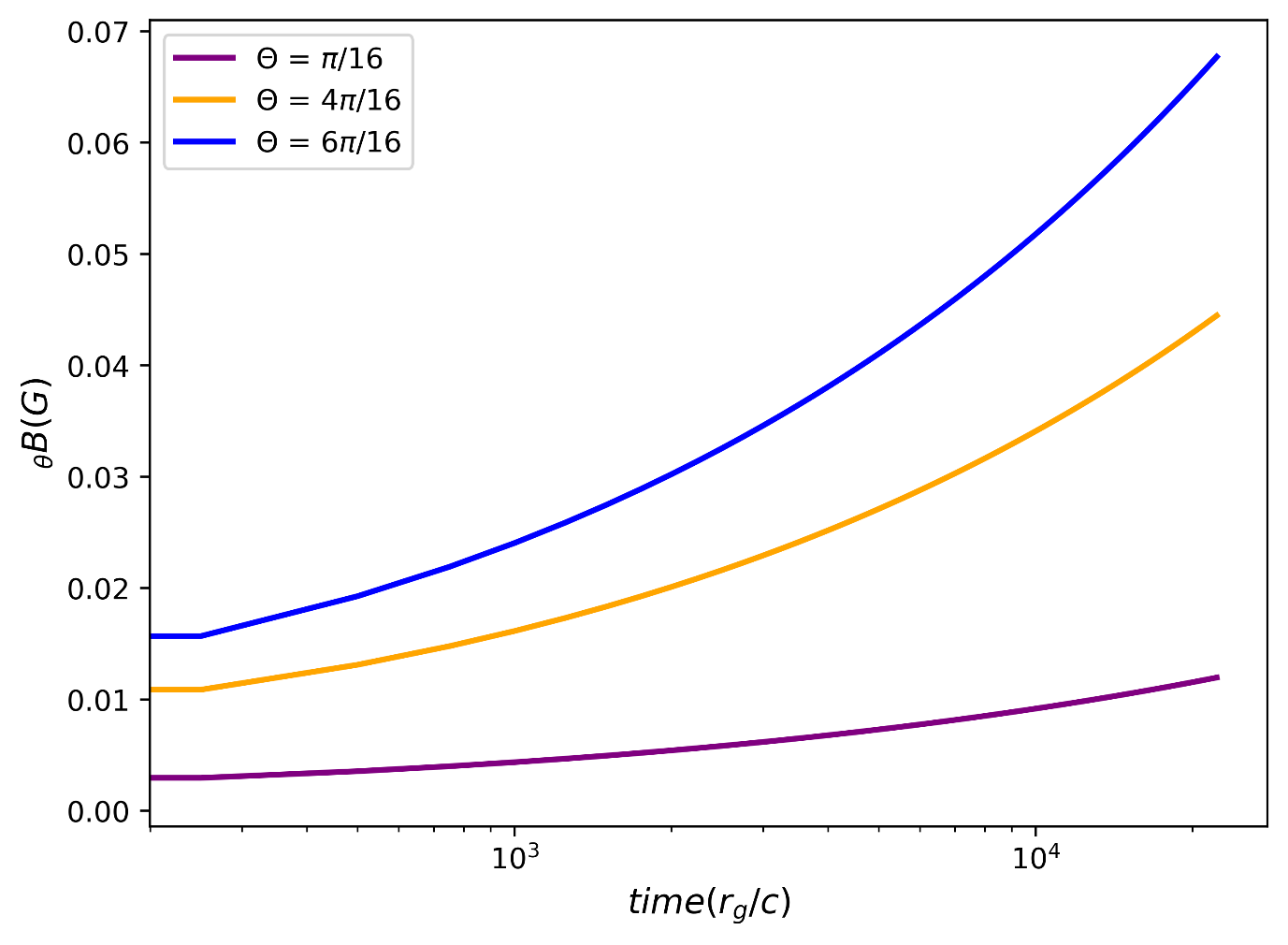}
        \caption{}
        \label{fig:fig2b}
    \end{subfigure}
    \caption{(a) Time evolution of the radial component of the magnetic field for Keplerian flow using Newton’s solution at different polar angles, (b) Time evolution of the polar component of the magnetic field for Keplerian flow using Newton’s solution at different polar angles.
}
    \label{fig:fig2}
\end{figure*}

We also compared the temporal evolution of the radial and polar physical components of the magnetic field in a Keplerian flow in Figure \ref{fig:fig2}. Figure \ref{fig:fig2a} relates to the radial component, while Figure \ref{fig:fig2b} relates to the polar component of the magnetic field. Each line in each graph corresponds to a different angle $\theta$ and shows how the physical components of the magnetic field change with time at different angles. This figure highlights the important differences in the temporal evolution of the physical components of the magnetic field. The radial $_r B$ component grows faster and at a higher rate than the $_\theta B $ component, with a very low value of $_\theta B $ indicating greater concentration of the magnetic field in the equatorial plane and its lower value in the polar regions. Additionally, the growth rate of each component varies with the angle $\theta $, with $_r B$ growing faster at smaller angles and $_\theta B $ at larger angles. These results emphasize the importance of considering angular dependence in the analysis of magnetic field evolution in accretion flows.

\subsection{Non-Keplerian flow equations}\label{sec:nonKepequ}

We consider the flow to be non-Keplerian and obtain the four-dimensional velocity components of the non-Keplerian flow from the integrals of motion, related to the conservation of energy $E$ and zero angular momentum $L = 0$. The angular velocity relation for the non-Keplerian flow is given by $\Omega = \frac{u^{\varphi }}{u^{t}} = \frac{(lg^{\varphi \varphi }-g^{t\varphi })}{(lg^{t\varphi }-g^{tt})}$. By expanding the relation $u^{i} u_{i}=1$ and substituting $u^\varphi$ from the relation $\Omega = \frac{u^\varphi }{u^t} = \frac{u_\varphi \frac{x^2}{r_{g}^2 \sin ^2\theta } }{\frac{1}{1-x} } = \frac{u_\varphi x^2\left(1-x\right) }{r_{g}^2\sin ^2\theta } $, we extract the radial component of the velocity in terms of the rotational component:
\begin{equation}    
    u^{r}=-\left\{x-u_{\varphi }^{2}  \frac{(1-x)x^2}{(r_{g} \sin \theta )^{2}} \right\}^{1/2}. 
    \label{eq:eq17}  
\end{equation} 
To separate it, we use the conservation of angular momentum:
\begin{equation}    
    L=\sum\limits_{k=1}^{3}{g_{ki} u^{i}=0}.
    \label{eq:eq18}
\end{equation}
Expanding equation~(\ref{eq:eq18}), and considering that $u^{\theta}=0$, we get:
\begin{equation}     
    g_{rr} u^{r}+g_{\varphi \varphi } u^{\varphi }=0.
    \label{eq:eq19}
\end{equation}
By integrating relations~(\ref{eq:eq17}) and~(\ref{eq:eq19}) and solving these equations simultaneously, we obtain $u^{r}$ and $u^{\varphi}$ components for non-Keplerian flow:
\begin{equation}
\begin{split}
    &u_{0}=1,\quad u^{0}=\left(1-x\right)^{-1}, \\
    &u^{r}=-\left\{\frac{x\left(1-x\right)\left(r_{g}\sin \theta \right)^{2}  }{x^{2}+\left(1-x\right)\left(r_{g}\sin \theta \right)^{2}} \right\}^{1/2},\\
    &u_{r}=\left\{\frac{x\left(r_{g}\sin \theta \right)^{2}  }{x^{2}\left(1-x\right) +\left(1-x\right)^{2}\left(r_{g}\sin \theta \right)^{2}} \right\}^{1/2}, \\
    & u^{\varphi} = \left\{\frac{x^{3} }{\left(1-x\right) \left(r_{g}\sin \theta \right) ^{2} \left[1+\frac{\left(1-x\right)\left(r_{g}\sin \theta \right) ^{2}  }{x^{2} } \right]  } \right\}^{1/2},\\
    &u_\varphi = \left\{\frac{x\left(r_{g}\sin \theta \right) ^{2} }{x^{2}\left(1-x\right)+\left(1-x\right) ^{2}\left(r_{g\sin \theta }\right) ^{2}    } \right\} ^{1/2 }. 
\end{split}
\label{eq:eq20}
\end{equation}

Finally, substituting the extracted components of the non-Keplerian flow velocity (relation~(\ref{eq:eq20})) into equation~(\ref{eq:eq7}) of the conservation of magnetic flux, we obtain the parametric components of the magnetic field for the non-Keplerian flow:
\begin{equation}
\begin{split}
    B^{\theta}=& -\frac{B_0}{r_g} \sin\theta \frac{x^{3/2 } }{\sqrt{x_{0} } }   \frac{\left(1-x_{0} \right) }{\left(1-x\right) } \left\{\frac{\left(1-x\right) \left(r_{g}\sin \theta  \right) ^{2} + x^{2}   }{\left(1-x_{0}\right) \left(r_{g}\sin \theta  \right) ^{2} + x_{0}^{2}} \right\} ^{1/2}, \\
    B^r =& \ B_0 \cos\theta \left(\frac{x}{x_0}\right)^2 \frac{(1-x)}{\sqrt{1-x_0}} \\
    &\left\{ \frac{x^2 + (1-x)(r_g \sin\theta)^2}{x^2 + (1-x)^2 (r_g \sin\theta)^2} \cdot \frac{x_0^2 + (1-x_0)^2 (r_g \sin\theta)^2}{x_0^2 + (1-x_0)(r_g \sin\theta)^2} \right\}.
\end{split}
\label{eq:eq21}
\end{equation}

\begin{figure*}
	\includegraphics[width=0.9\textwidth]{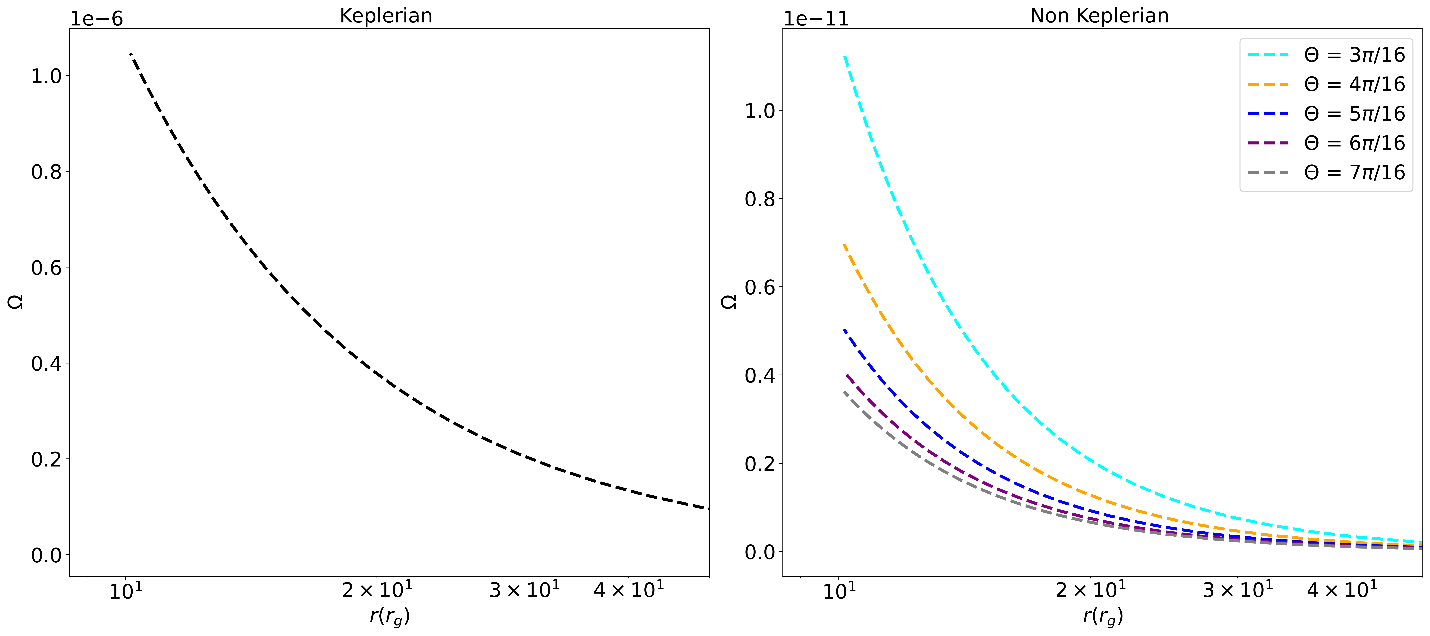}
    \caption{The behavior of the angular velocity component for Keplerian and non-Keplerian flows as a function of radius.
    \label{fig:fig3}}
\end{figure*}

Figure \ref{fig:fig3} shows the angular velocity $\Omega $ as a function of $r$ for Keplerian and non-Keplerian flows, allowing us to determine that the non-Keplerian flow here is a sub-Keplerian flow. So from here on we compare two types of Keplerian and sub-Keplerian flow.

\begin{figure*}
	\includegraphics[width=0.9\textwidth]{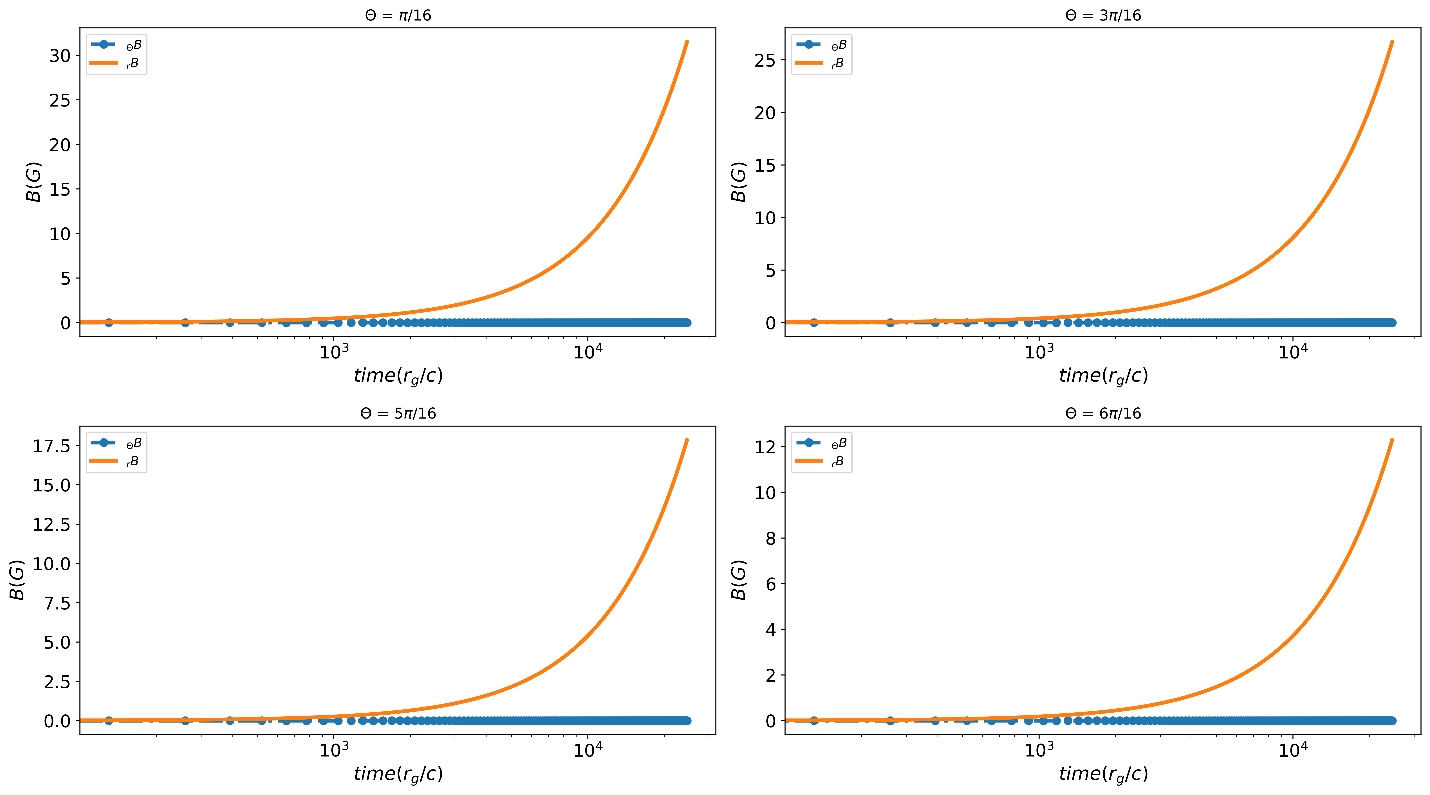}
    \caption{Comparison of the time evolution of the physical components of the magnetic field in sub-Keplerian flow at different polar angles. 
    \label{fig:fig4}}
\end{figure*}

In Figure \ref{fig:fig4}, we plot the time evolution of the physical components of the magnetic field in a sub-Keplerian flow at four angles $\theta =\frac{\pi }{16}, \frac{3\pi }{16}, \frac{5\pi }{16}, \frac{6\pi }{16} $. Following \citeauthor{2019Univ....5..146B}'s method, we selected this range of angles to cover important physical regions and dynamic processes, such as magnetic interactions and accretion of matter near the equatorial plane. This range is determined by excluding two cones with opening angles of $\frac{\pi}{8}$, resulting in the polar angle being confined to $\frac{\pi}{16}$ to $\frac{15\pi}{16}$ \citep{2019Univ....5..146B}. This approach reduces boundary issues and improves numerical stability. 
The diagrams show how the physical components of the magnetic field change with time for different angles. Figure \ref{fig:fig4}, demonstrates that as the angle $\theta $ increases, there is a continuous pattern in the growth of $_r B$ and the relative stability of $_\theta B $. The magnitude of $_r B$ increases more rapidly for larger angles, while $_\theta B $ remains relatively stable with minimal changes.

\begin{figure*}
    \centering
    \begin{subfigure}{0.45\textwidth}
        \includegraphics[width=\textwidth]{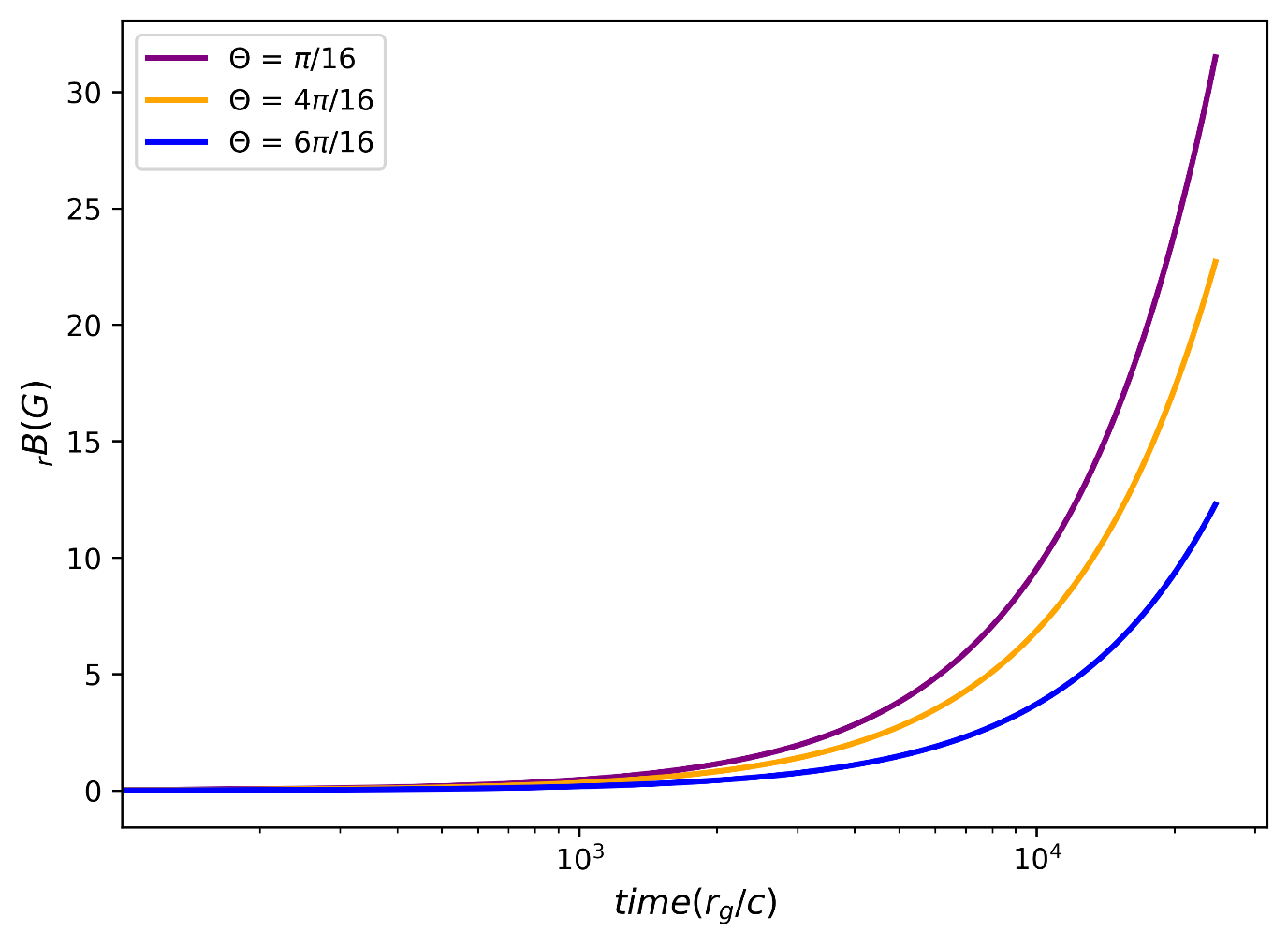}
        \caption{}
        \label{fig:fig5a}
    \end{subfigure}
    \hspace{-0.01\textwidth} 
    \begin{subfigure}{0.45\textwidth}
        \includegraphics[width=\textwidth]{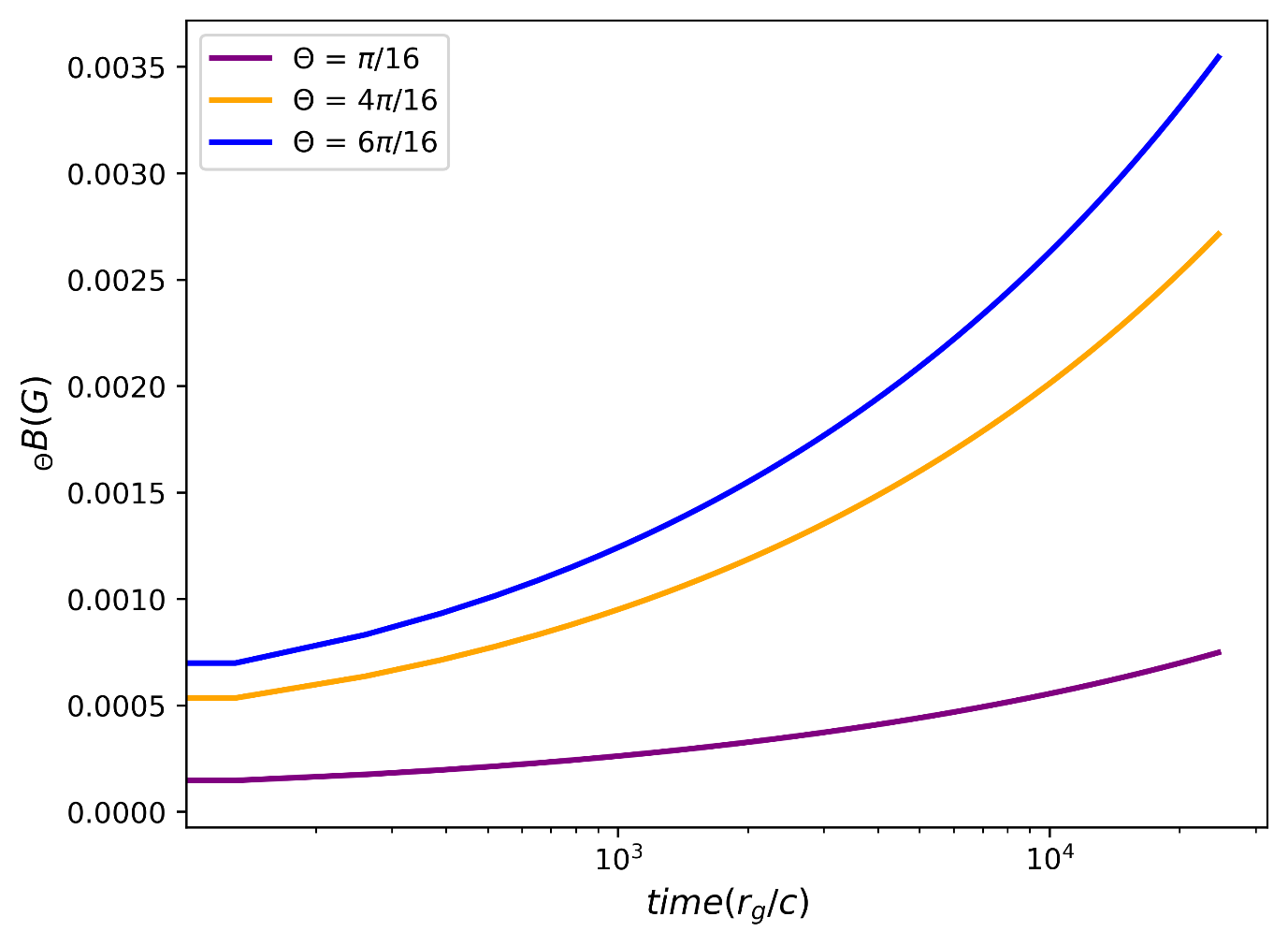}
        \caption{}
        \label{fig:fig5b}
    \end{subfigure}
    \caption{(a) Time evolution of the radial component of the magnetic field for sub-Keplerian flow using Newton’s solution at different polar angles, (b) Time evolution of the polar component of the magnetic field for sub-Keplerian flow using Newton’s solution at different polar angles.}
    \label{fig:fig5}
\end{figure*}

Additionally, Figure \ref{fig:fig5} presents the temporal evolution of the radial and polar physical components of the magnetic field in a sub-Keplerian flow. Figure \ref{fig:fig5a} relates to the radial component, and Figure \ref{fig:fig5b} to the polar component. Each line in these diagrams corresponds to a different angle $\theta$, showing important differences in the temporal evolution of these magnetic field components. The radial $_r B$ component grows faster compared to the polar component $_\theta B $. The exponential growth of $_r B$ indicates a strong field amplification mechanism in the radial direction. This can be attributed to the differential rotation of the accretion flow, where the radial shear generates and amplifies the magnetic field and very low value of $_\theta B $ indicates the greater concentration of the magnetic field in the equatorial plane and its low value in the polar regions In addition, the growth rate of each component varies with the angle $\theta $. So that the growth speed of $_r B$ is at smaller angles and $_\theta B $ is at larger angles. This time dependence suggests that the radial magnetic field becomes increasingly dominant as the accretion flow evolves and plays a significant role in the dynamics of the accretion process. However, considering the angular dependence is crucial in analyzing the evolution of the magnetic field in accretion flows.

\section{PARAMETERIZATION OF EQUATION}\label{sec:paraequ}
In contrast to the specific case considered by \citeauthor{2019Univ....5..146B}, where $r< r_c$ and the flow near the BH is assumed to be in free-fall, our work generalizes this scenario by considering a broader range of flow conditions. \citeauthor{2019Univ....5..146B}'s analysis was limited to free-fall conditions close to the BH, which, while insightful, does not encompass the full spectrum of potential flow behaviors. By expanding the scope to include both Keplerian and sub-Keplerian flows, we provide a more general framework for understanding the evolution of magnetic fields in these environments.
In this section, we aim to parametrize the equations in the $\left(r> r_c\right) $ regim. First, it is necessary to clarify what we mean by $r> r_c$. Here, we refer to the zone where the flow is subsonic and located in distant regions, where relativistic effects are negligible, and the Newtonian limit holds. This implies that the Newtonian limit $\left(0<x,\ x_{0}\ll 1\right) $ is applicable in this range as $x\rightarrow 0$, which we will apply to our relations. To investigate the evolution of the magnetic field components over time, we first obtain the integral response of equation~(\ref{eq:eq10}) for both Keplerian and sub-Keplerian flows. By solving equation~(\ref{eq:eq10}), the time $t$ enters the integrals explicitly, so the magnetic field at a given spatial point depends on time in a hydrodynamically stationary flow. Then, we apply the Newtonian limit to derive the dependence of $x$ in terms of $t$. In this way, the components of the magnetic field have an implicit dependence on $t$.

\subsection{Keplerian flow}\label{sec:kepflow}
By substituting the components of the Keplerian flow velocity into equation~(\ref{eq:eq10}), we derive the following integral relation, which we solve numerically. However, since we aim to express the function $t$ in terms of $x$, it is necessary to fit an appropriate function:
\begin{equation}
    \frac{ct}{r_g} + \int \frac{dx}{x^2 (1-x) \left[ x - \frac{x \sin^2 \theta}{(1-x)} \right]^{1/2}} = c_1.
    \label{eq:eq22}
\end{equation} 
By integrating equations~(\ref{eq:eq9}) and~(\ref{eq:eq10}) with relation~(\ref{eq:eq13}) and applying  the initial condition~(\ref{eq:eq15}) in a parametric form, we obtain a solution that describes the magnetic field for the Keplerian flow:
\begin{equation}
    \frac{ct}{r_g} + Bx + C x^{3/2} + D x^{-3/2} + E x^{2} = Bx_{0} + C x_{0}^{3/2} + D x_{0}^{-3/2} + E x_{0}^{2},
    \label{eq:eq23}
\end{equation}
where $B=21.96875$, $C=4.1551045$, $D=1.0997513$ and $E=0.012491225$.
We then derive the dependence of $x$ in terms of $t$ for the Keplerian flow:
\begin{equation}
    x_0 = x \left(1 + \frac{ct x^{3/2}}{D r_g}\right)^{-2/3}.   
    \label{eq:eq24}
\end{equation}
Next, we obtain the components $B^{r}$ and $B^{\theta}$ of the magnetic field for the Keplerian flow by substituting $x_{0}$ into equation~(\ref{eq:eq16}): 
\begin{equation}
\begin{split}
    B^{\theta}=& -\frac{B_0}{r_g} \sin\theta    x\left(1-x\right)  \left(1+ \frac{ctx^{3/2 } }{D r_{g} } \right) ^{2/3 }     \\
    &      \left\{\frac{  \left(1-x\left(1+ \frac{ctx^{3/2 } }{D r_{g} } \right) ^{2/3 }\right)-\sin ^{2}\theta    }{   \left(1+ \frac{ctx^{3/2 } }{D r_{g} } \right) ^{2/3 }    \left(\left(1-x\right)-\sin ^{2} \theta\right)} \right\} ^{1/2 },  \\
    B^r =& \ B_0 \cos\theta     \left( 1-x \right)^2   \left( 1 + \ \frac{ct x^{3/2}}{D r_g} \right)^{4/3}       \\
    &   \left\{ \frac{\left( 1 - x \left( 1 +  \frac{ct x^{3/2}}{D r_g} \right)^{-2/3} \right)^{2}  +  x \left( 1 +  \frac{ct x^{3/2}}{D r_g} \right)^{-2/3}\sin^2 \theta }     
    {    \left[ 1 - x \left( 1 +  \frac{ct x^{3/2}}{D r_g} \right)^{-2/3} \right]^{3/2}    \left( \left( 1 - x \right)^{2}  +x \sin^2 \theta\right) } \right\}.  
\end{split} 
\label{eq:eq25}
\end{equation}    

\begin{figure*}
	\includegraphics[width=0.9\textwidth]{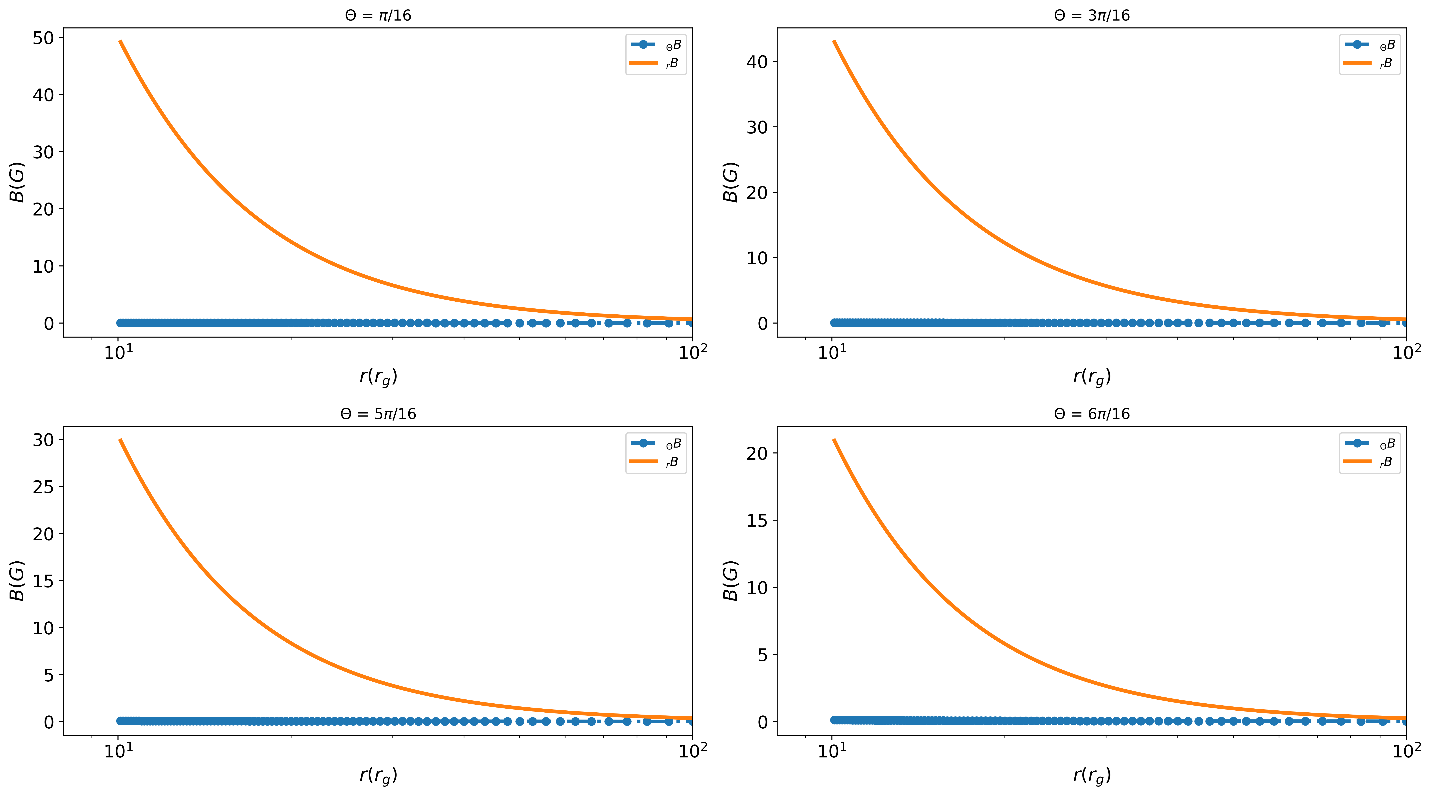}
    \caption{Comparison of the variations in the radial and polar components of the Keplerian flow magnetic field as a function of radius for different polar angles. 
    \label{fig:fig6}}
\end{figure*}

Using the above calculations, we plotted the behavior of the physical components of the magnetic field for the Keplerian flow as a function of radius. In Figure \ref{fig:fig6}, the radial $\left(_r B\right) $ and polar $\left(_\theta B\right) $ components of the magnetic field are shown for different angles $\theta =\frac{\pi }{16}, \frac{3\pi }{16}, \frac{5\pi }{16}, \frac{6\pi }{16} $. Each plot corresponds to a different angle$\left(\theta\right)$ and illustrates how the magnetic field components change with radius. We observe that $_r B$ has higher values at smaller radii and decreases as the radius increases. On the other hand, $_\theta B$ has much lower values compared to $_r B$ and shows less variation with angle, though it gradually decreases with increasing radius.

\begin{figure*}
    \centering
    \begin{subfigure}{0.45\textwidth}
        \includegraphics[width=\textwidth]{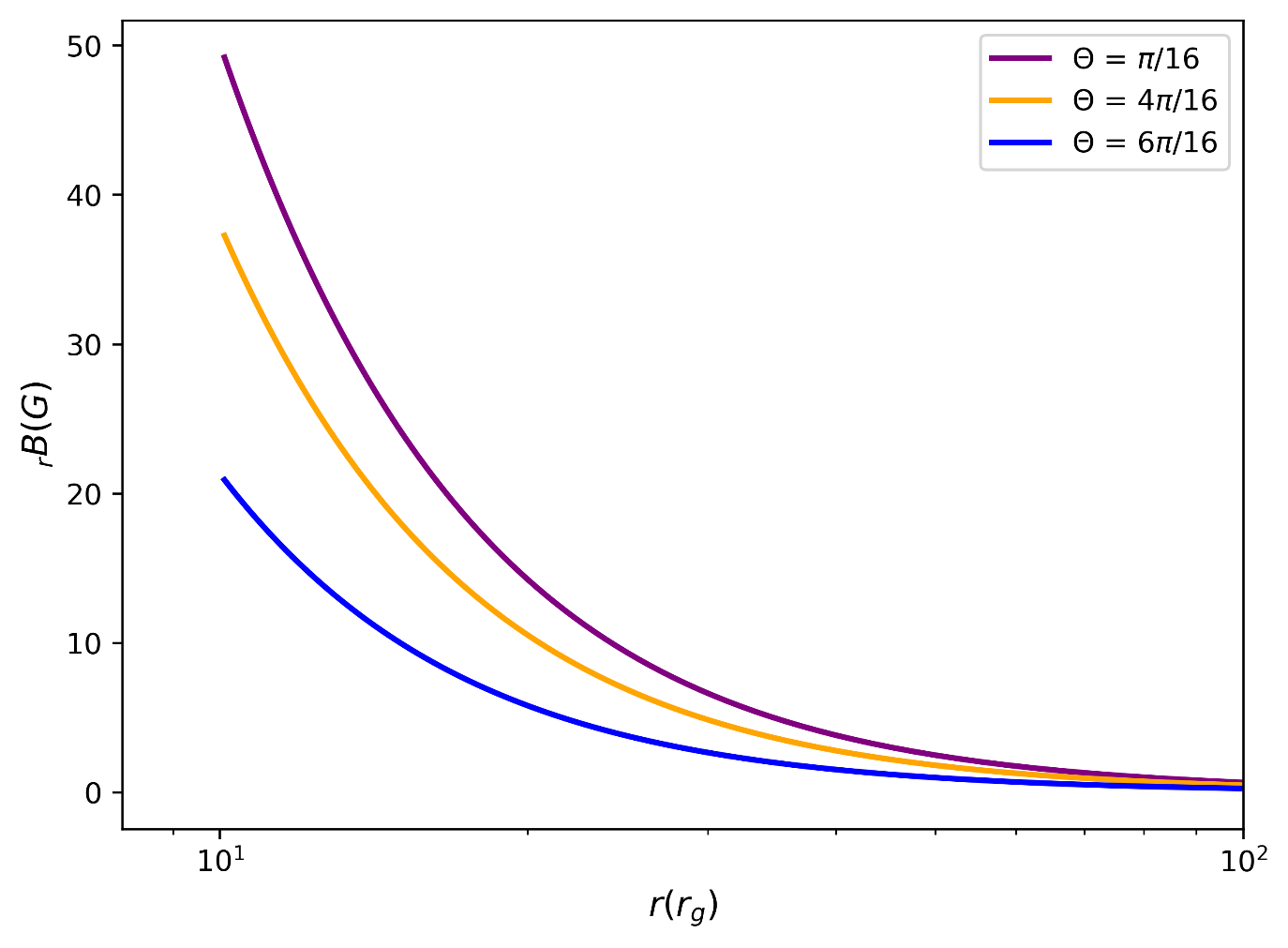}
        \caption{}
        \label{fig:fig7a}
    \end{subfigure}
    \hspace{-0.01\textwidth} 
    \begin{subfigure}{0.45\textwidth}
        \includegraphics[width=\textwidth]{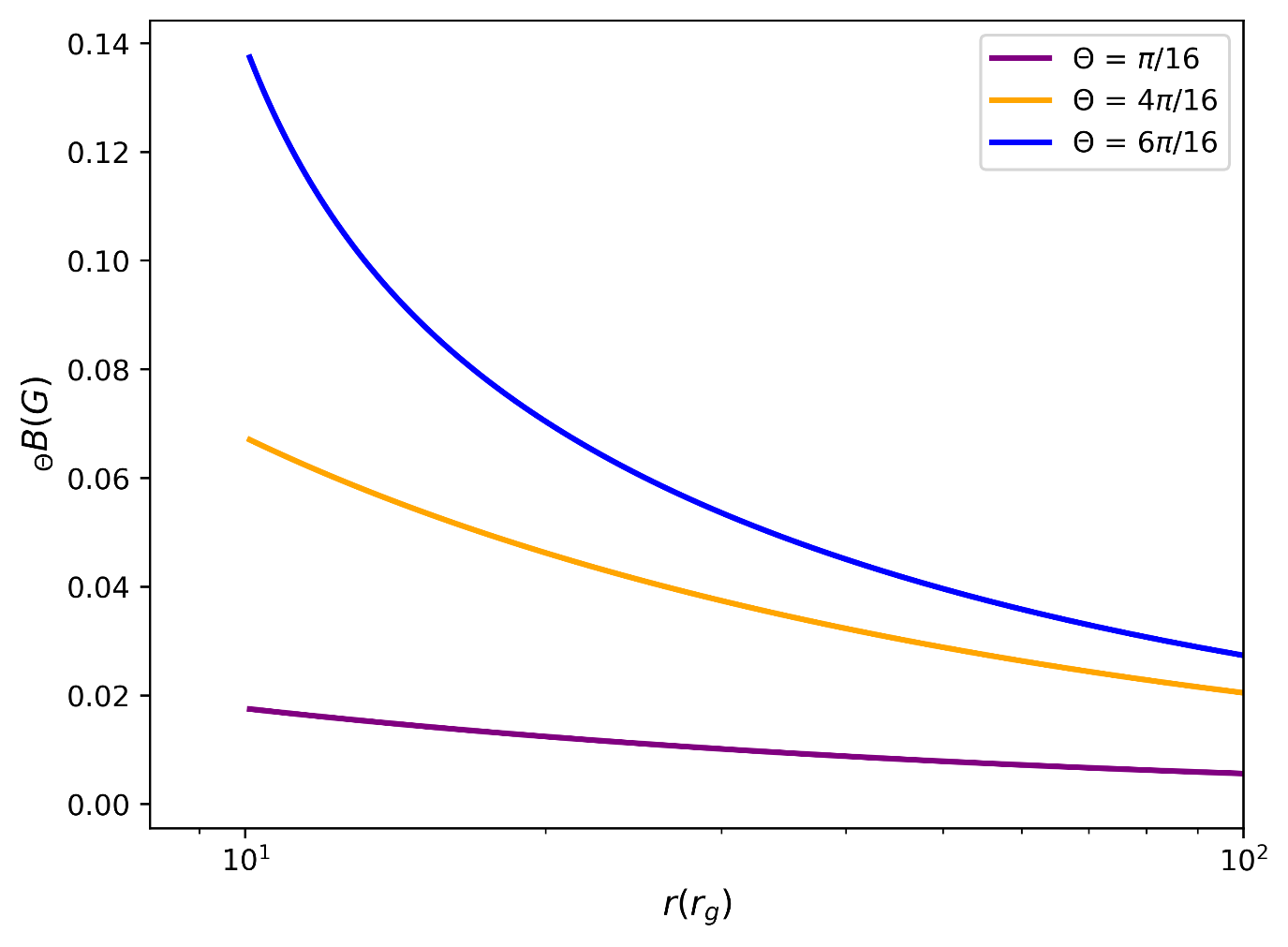}
        \caption{}
        \label{fig:fig7b}
    \end{subfigure}
    \caption{(a) The variations in the radial component of the Keplerian flow magnetic field as a function of radius for different polar angles, (b) The variations in the polar component of the Keplerian flow magnetic field as a function of radius for different polar angles.
}
    \label{fig:fig7}
\end{figure*}

In addition, Figure \ref{fig:fig7}, compares the changes in the physical components of the magnetic field for the Keplerian flow as a function of radius. Figure \ref{fig:fig7a} illustrates the variations in the $_r B$ component at three different angles with respect to radius, while Figure \ref{fig:fig7b} shows the $_\theta B$ component at the same angles. From Figure \ref{fig:fig7a}, we see that the radial component of the magnetic field decreases with increasing radius, indicating that the magnetic field is strongest near the BH and weakens as the distance from the center increases. Figure \ref{fig:fig7b} illustrates that the polar component also decreases with increasing radius, but to a lesser extent than the radial component. The low value of $_\theta B$ suggests that the magnetic field is concentrated in the equatorial plane and is weaker in the polar regions.

\subsection{Sub-Keplerian flow}\label{sec:nonkepflow}
By substituting the components of the sub-Keplerian flow velocity into relation~(\ref{eq:eq10}) for the sub-Keplerian flow, we derive the following integral relation:
\begin{equation}
    \frac{ct}{r_g} - \int \frac{\sqrt{(1-x) (r_g \sin\theta)^2 + x^2}}{x^2 \sqrt{x(1-x)^3 (r_g \sin\theta)^2}} \, dx = c_1.
    \label{eq:eq26}
\end{equation}
By integrating equations~(\ref{eq:eq9}) and~(\ref{eq:eq10}) with relation~(\ref{eq:eq20}) and replacing the initial condition~(\ref{eq:eq15}) in a parametric form, we obtain a solution that defines the magnetic field of sub-Keplerian flow:
\begin{equation}
    \frac{ct}{r_g} + \frac{2}{3} x^{-3/2} + 2 x^{-1/2} + \ln \left( \frac{1 - \sqrt{x}}{1 + \sqrt{x}} \right) = \frac{2}{3} x_0^{-3/2} + 2 x_0^{-1/2} + \ln \left( \frac{1 - \sqrt{x_0}}{1 + \sqrt{x_0}} \right).
    \label{eq:eq27}
\end{equation}
We obtain the dependence of $x$ in terms of $t$ for the sub-Keplerian flow as:
\begin{equation}
    x_0 = x \left( 1 + \frac{3}{2} \frac{ct x^{3/2}}{r_g} \right)^{-2/3}.
    \label{eq:eq28}
\end{equation}
We obtain the components $B^{r}$ and $B^{\theta}$ of the field for the sub-Keplerian flow by inserting $x_{0}$ into equation~(\ref{eq:eq20}):
\begin{equation}
\begin{split}
    B^{\theta} =& -\frac{B_0}{r_g} \sin \theta \frac{x}{1-x}   \left(1+\frac{3}{2} \frac{ctx^\frac{3}{2}  }{r_{g} } \right) ^\frac{1}{3}    \left[1-x\left(1+\frac{3}{2} \frac{ctx^\frac{3}{2}  }{r_{g} } \right) ^\frac{-2}{3}  \right] ^\frac{1}{2}    \\  
    &   \left\{\frac{\left(1-x\right)\left(r_{g}\sin \theta  \right) ^{2}+x^{2}   }{\left[1-x\left(1+\frac{3}{2} \frac{ctx^\frac{3}{2}  }{r_{g} } \right) ^\frac{-2}{3}  \right]\left(r_{g}\sin \theta  \right) ^{2}+x^{2}\left(1+\frac{3}{2} \frac{ctx^\frac{3}{2}  }{r_{g} } \right) ^\frac{-4}{3} } \right\} ^\frac{1}{2}   ,\\
    B^r =& \ B_{0} \cos \theta   \left(1-x\right)    \left(1+\frac{3}{2} \frac{ctx^\frac{3}{2}  }{r_{g} } \right) ^\frac{4}{3}   \\
    &   \left[1-x\left(1+\frac{3}{2} \frac{ctx^\frac{3}{2}  }{r_{g} } \right) ^\frac{-2}{3}  \right] ^\frac{-1}{2}    \left\{\frac{\left(1-x\right)\left(r_{g}\sin \theta  \right) ^{2}+x^{2}}{\left(1-x\right)^{2} \left(r_{g}\sin \theta  \right) ^{2}+x^{2}}   \right\}  \\
    &    \left\{ \frac{\left[1-x\left(1+\frac{3}{2} \frac{ctx^\frac{3}{2}  }{r_{g} } \right) ^\frac{-2}{3}  \right]^{2} \left(r_{g}\sin \theta  \right) ^{2}+x^{2}\left(1+\frac{3}{2} \frac{ctx^\frac{3}{2}  }{r_{g} } \right) ^\frac{-4}{3}   }{\left[1-x\left(1+\frac{3}{2} \frac{ctx^\frac{3}{2}  }{r_{g} } \right) ^\frac{-2}{3} \right]\left(r_{g}\sin \theta  \right) ^{2}+x^{2}\left(1+\frac{3}{2} \frac{ctx^\frac{3}{2}  }{r_{g} } \right) ^\frac{-4}{3} } \right\}  .
\end{split} 
\label{eq:eq29}     
\end{equation} 
 
\begin{figure*}
	\includegraphics[width=0.9\textwidth]{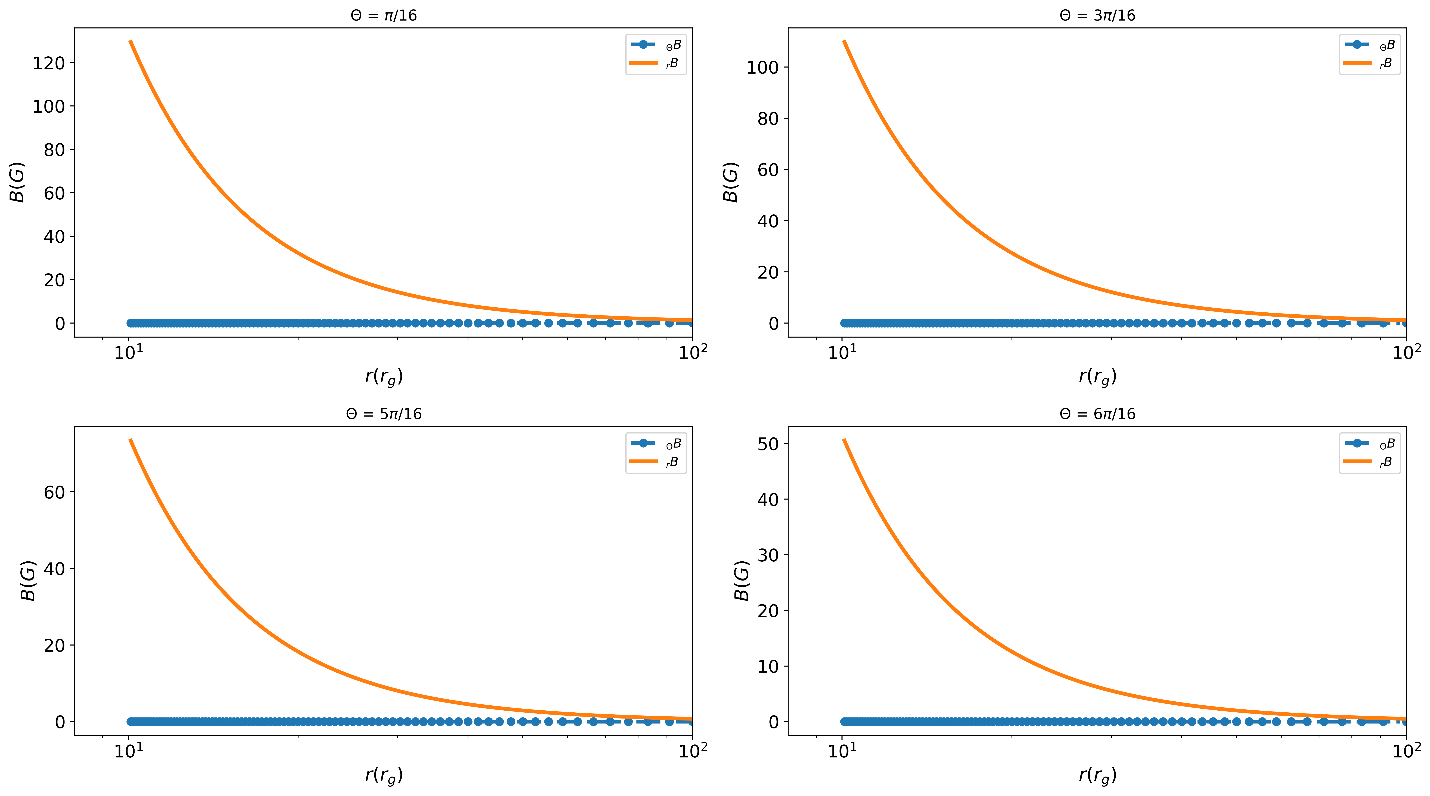}
    \caption{ Comparison of the variations in the radial and polar components of the sub-Keplerian flow magnetic field as a function of radius for different polar angles. 
    \label{fig:fig8}}
\end{figure*}

For the sub-Keplerian flow, considering the above calculations, we plotted the behavior of the physical components of the magnetic field as a function of the radius. In Figure \ref{fig:fig8}, we plotted the changes in the radial $\left(_r B\right) $ and polar $\left(_\theta B\right) $ components of the magnetic field as a function of the radius for the sub-Keplerian flow at four different angles $\theta =\frac{\pi }{16}, \frac{3\pi }{16}, \frac{5\pi }{16}, \frac{6\pi }{16} $. Each plot in each figure corresponds to a different angle $\left(\theta\right)$ and shows how the magnetic field components change with radius. In Figure \ref{fig:fig8}, we see that the radial and polar components of the magnetic field in the sub-Keplerian flow decrease with increasing radius. The radial component $_r B$ has a larger value at all angles and decreases faster, while the polar component $_\theta B$ gradually decreases with increasing radius and does not change much.
   
\begin{figure*}
    \centering
    \begin{subfigure}{0.45\textwidth}
        \includegraphics[width=\textwidth]{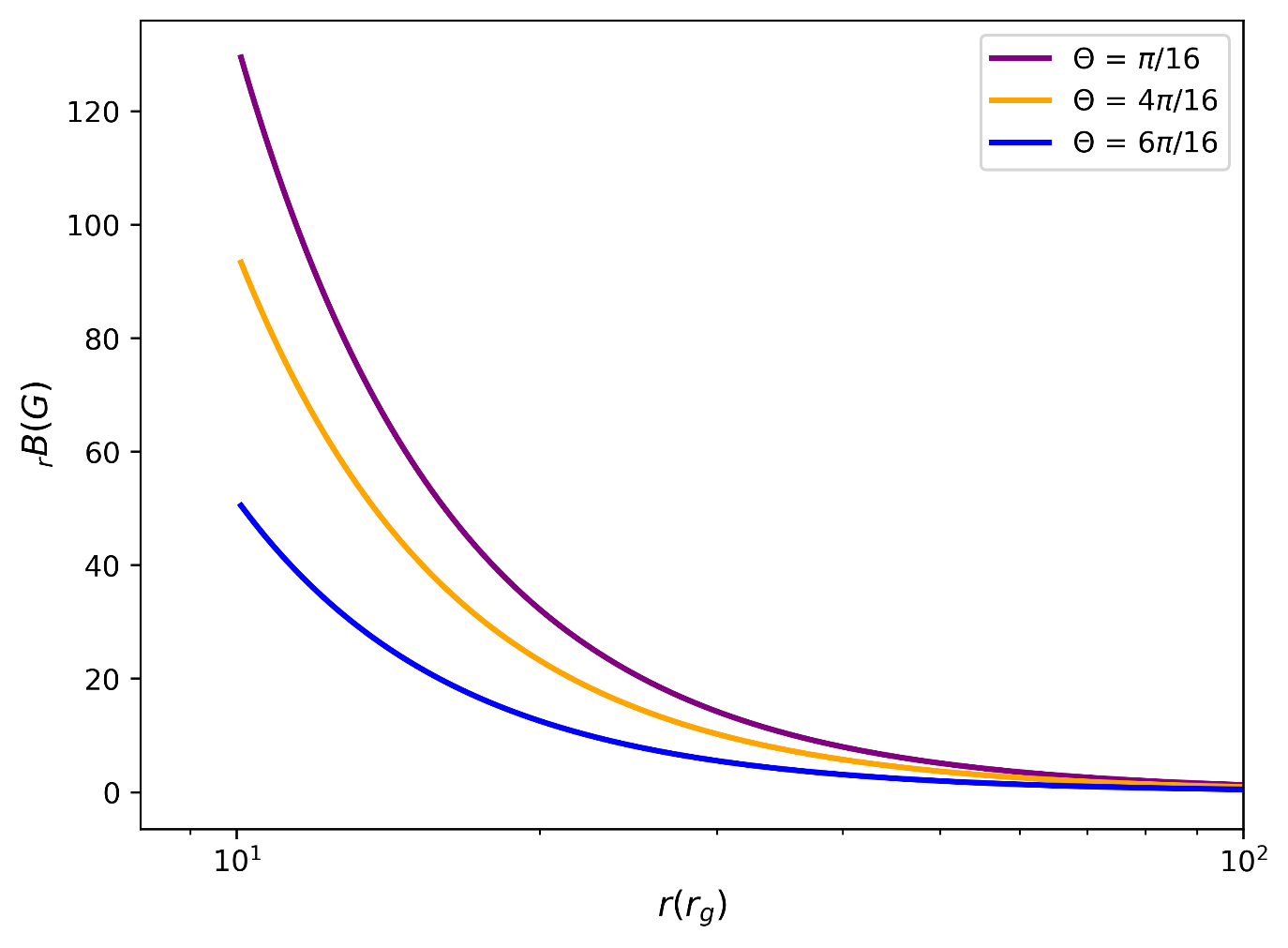}
        \caption{}
        \label{fig:fig9a}
    \end{subfigure}
    \hspace{-0.01\textwidth} 
    \begin{subfigure}{0.45\textwidth}
        \includegraphics[width=\textwidth]{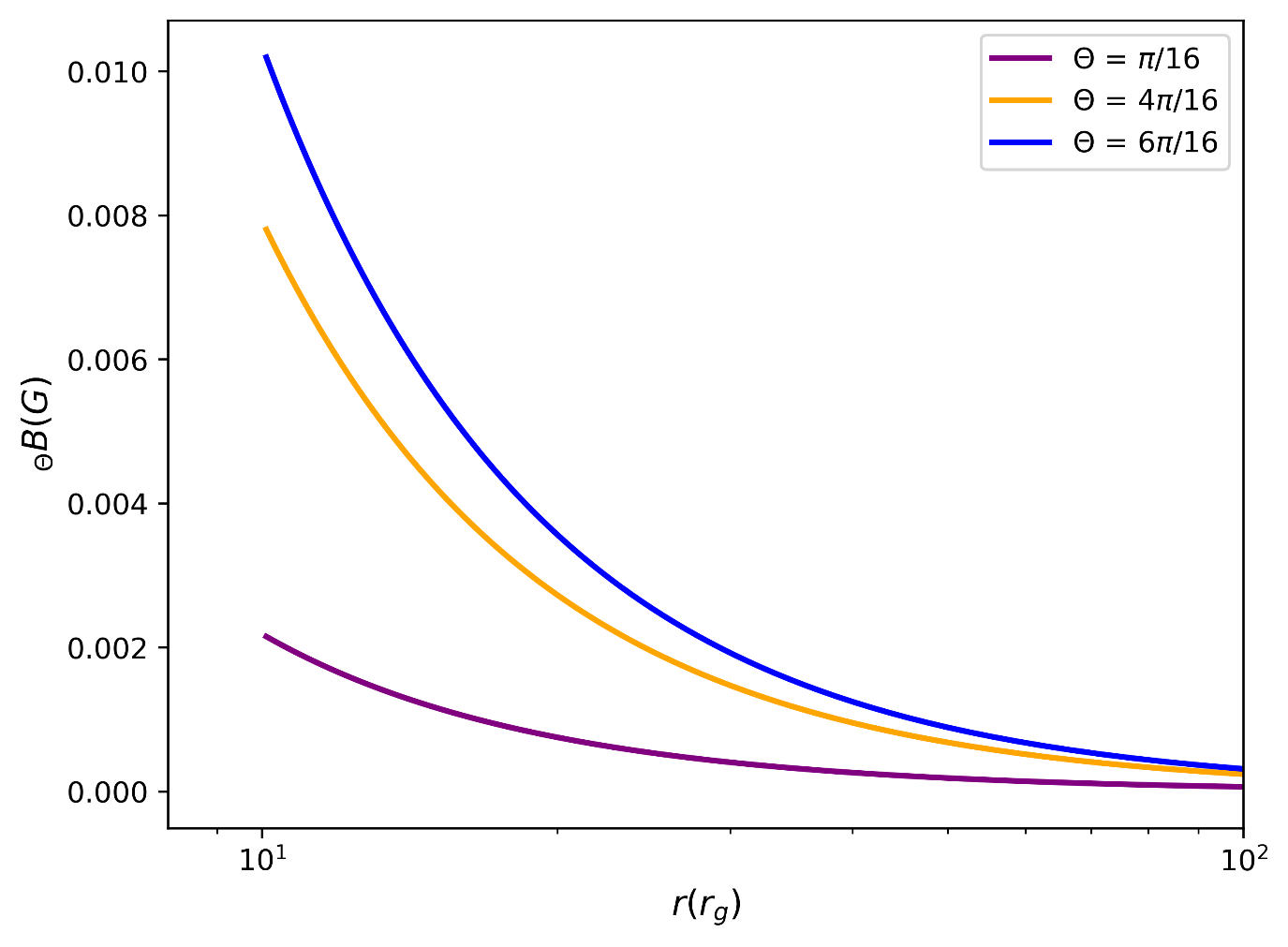}
        \caption{}
        \label{fig:fig9b}
    \end{subfigure}
    \caption{(a) Comparison of the variations in the radial component of the sub-Keplerian flow magnetic field as a function of radius for different polar angles, (b) Comparison of the variations in the polar component of the sub-Keplerian flow magnetic field as a function of radius for different polar angles.
}
    \label{fig:fig9}
\end{figure*}

In Figure \ref{fig:fig9}, we plotted and compared the changes in the physical components of the magnetic field for the sub-Keplerian flow as a function of radius. Figure \ref{fig:fig9a} shows the changes in the $_r B$ component of the field at three different angles with respect to the radius, and Figure \ref{fig:fig9b} shows the $_\theta B$ component of the magnetic field at the same three angles as a function of the radius.

In Figure \ref{fig:fig9a}, we see that the radial component of the magnetic field in the sub-Keplerian flow decreases with increasing radius. This shows that the maximum strength of the magnetic field is near the BH, and the strength of the magnetic field decreases as you move away from the center of the BH. According to Gauss's law, the inverse square law, and the law of the dynamo effect, we expect such a decreasing behavior of the physical components of the magnetic field with distance from the BH. Our results are consistent with these three physical laws, and gradually with increasing radius, the physical radial component of the magnetic field becomes weaker. In Figure \ref{fig:fig9b}, the polar component of the magnetic field also decreases with increasing radius and distance from the BH, though it diminishes to a lesser extent than the radial component. A very low value of $_\theta B$ indicates the concentration of the magnetic field in the equatorial plane, with this component being less prominent in the polar regions. 

\section{DISCUSSION AND CONCLUSION }\label{sec:conclusion}
Accretion discs are capable of hosting magnetic fields of both small and large scales. At the outset, a fundamental question arises: What could be the potential source of both small and large-scale magnetic fields within accretion discs? Astrophysicists have extensively investigated this issue in their research. They confirm that small-scale magnetic fields can be locally generated by turbulence resulting from rotational magnetic instabilities. This type of magnetic field can contribute to the transfer of angular momentum away from the accretion disc. Nevertheless, the process of generating large-scale magnetic fields within accretion discs remains a subject of ongoing investigation, and astrophysicists have proposed various hypotheses to elucidate this phenomenon, a selection of which we will discuss below. 
The second question that arises pertains to whether there exists an upper limit to the amount of magnetic flux capable of arresting the accretion flow. The upper limit under consideration for the magnetic field surrounding BHs is referred to as the Eddington magnetic field. This limit is determined by comparing the energy density of the magnetic field with that of the accreting plasma responsible for the Eddington luminosity near a BH, and its value is calculated as $B_{Edd}  \approx  10^4 G$ \citep{Beskin2009}. An approximate upper limit for the amount of magnetic field strength that any disc around a BH can sustain may be the Eddington limit of the magnetic field near the event horizon of a BH \citep{2018MNRAS.476.2396M}. Based on several observational and theoretical models, values of the magnetic field near the event horizon of stellar-mass BHs are proposed to be $B \sim  10^8  G$, while for massive BHs, $B \sim  10^4  G$ \citep{2011AstBu..66..320P}. 

In the model we have examined, the initial assumption is that the magnetic field originates from the outer environment of the disc and enters the accretion environment by adhering to the flow. During the accretion process, magnetized flow with initially weak magnetic fields is drawn toward a BH. Over time, as these flows approach the vicinity of a BH, significant compression of the magnetic field occurs in its proximity. This compression arises due to the BH's gravitational attraction, which draws in plasma while resisting the inclusion of magnetic field lines. Furthermore, the continuous inward accretion of magnetized flux within the quasi-spherical accretion flow contributes to the ongoing evolution and amplification of this large-scale magnetic field. Gradually, the magnetic field's magnitude increases to a point where it attains dynamic dominance, effectively serving as an impediment to the advancing flow and decelerating its motion. We have investigated the behavior of this magnetic field considering both Keplerian and sub-Keplerian flows, utilizing Newton's solution in regions distant from a BH. The question that arises here is whether the large-scale magnetic field can attain its maximum strength in the distant region or not. Based on the obtained results, it can be observed that, in general, the radial component of the magnetic field exhibits a faster growth rate compared to its polar component. This trend is observed for both flows. In our model, the zone in areas distant from a BH spans $r =20 r_{g}$ in radius. Additionally, we considered the initial field that is drawn to the Newtonian zone by the accretion plasma to be $10 mG$.

During accretion of the matter with a large scale magnetic field into a BH, a disc is formed around a BH, which equilibrium supported by the balance between BH gravity and a magnetic pressure. The polar domain extends from $\theta  = {\pi }/{16}$ to $\theta  = {15\pi }/{16}$. The field strength in the polar direction gradually changes from low intensity around the equatorial plane to higher intensity near the poles.

Our results in Figures \ref{fig:fig1}, \ref{fig:fig2}, \ref{fig:fig4} and \ref{fig:fig5} show the time evolution of the physical components of the magnetic field for Keplerian and sub-Keplerian flows, respectively. In both types of flows, a similar behavior is observed between the radial and polar components of the magnetic field over time. The reason for choosing a time scale for magnetic field evolution up to order $10^4$ is that it is consistent with previous studies on the time scale of magnetic field evolution, including the work by Avara et al. (2016). In this time scale, physical processes such as magnetic dynamos and the inward transfer of magnetic flux are significantly activated \citep{10.1093/mnras/stw1643}. 
Within this time scale, dynamo effects—caused by the differential movements of the accretion flow—are strengthened, leading to a significant increase in the components of the magnetic field, particularly the radial component $\left(_r B\right) $. This timescale is also long enough to observe gradual changes and long-term dynamics in the magnetic field, while still being shorter than the timescales associated with larger galactic-scale dynamics. As a result, a timescale up to the order of $10^4$ has been chosen to accurately display the evolution and strengthening of the magnetic field in accretion flows and to depict the complex and long-term effects of magnetic and gravitational interactions around BHs.

In plots of magnetic field evolution, we observe that the exponential growth of $_r B$ indicates a strong amplification mechanism in the radial direction, which can be attributed to the differential rotation of the accretion flow, where radial shear generates and amplifies the magnetic field. The time dependence shows that as the accretion flow evolves, the radial magnetic field becomes increasingly dominant and can play an important role in the dynamics of the accretion process. This predominance could be due to the complex interactions between the BH's gravitational force and magnetic fields, as well as dynamo effects that are amplified near the BH, thereby enhancing the radial magnetic field.
The relatively stable behavior of $_\theta B$ indicates that the polar component of the magnetic field is less affected by the dynamical processes in the accretion flow. This stability may result from the fact that dynamic changes in these regions are mainly driven by the distribution of the magnetic field in the radial direction, with less impact on the polar component. This suggests that the driving mechanisms of magnetic field evolution are primarily concentrated in the radial direction and near the BH. By plotting at different angles, we were able to reveal the angular effects on the growth of the field components. The dependence of $_r B$ on $\theta$ shows that the inclination angle plays a role in the efficiency of magnetic field amplification. For larger angles, the flow geometry may amplify shear effects, leading to a faster increase in the radial component of the magnetic field. This may be due to changes in the flow structure and distribution of the magnetic field with angle, which causes the magnetic field to strengthen and expand more rapidly at these angles. On the other hand, the dependence of $_\theta B$ on $\theta$ suggests that in regions where $_\theta B$ is low, the magnetic field is more concentrated in the equatorial plane. This is due to the greater distance from the field source in these regions, as well as the rotational and shear effects that disperse the field in the equatorial plane. In areas where $_\theta B$ is high, the magnetic field is strengthened and more concentrated near the poles due to the concentration of field lines in the polar regions and dynamo effects, which indicates a greater concentration of the magnetic field in the polar regions.
According to Figures \ref{fig:fig1} and \ref{fig:fig4}, since $_r B$ grows faster and at a higher rate than $_\theta B$ and the changes in $_\theta B$ are minimal, it can be concluded that the evolution of the magnetic field is quasi-radial.

As shown in Figure \ref{fig:fig1}, over time, the magnitude of the radial component of the field for the Keplerian flow in regions away from the BH around $\theta  = {\pi }/{16}$ reaches $12 G$. Similarly, in Figure \ref{fig:fig4}, the magnitude of the radial component of the magnetic field in the same region is $30 G$ for the sub-Keplerian flow. In some sources and simulations of this model, it appears that the magnitude of the magnetic field in the region around the BH $r =20 r_{g}$ is quantitatively confirmed to be within the range $1G \leq B \leq 30G$, which is consistent with our results in this paper \citep{Akiyama_2021}.

Our results in Figures \ref{fig:fig6}, \ref{fig:fig7}, \ref{fig:fig8} and \ref{fig:fig9} show the behavior of the physical components of the magnetic field as a function of radius for Keplerian and sub-Keplerian flows, respectively. In both types of flows, similar behavior is observed between the radial and polar components of the magnetic field as the radius changes. These plots demonstrate that the maximum strength of the magnetic field is near the BH, and the strength decreases as one moves away from the BH's center. According to Gauss's laws, the inverse square law, and the law of the dynamo effect, such a decreasing behavior of the magnetic field components with the distance from the BH is expected. Generally, the magnetic field in space decreases as the inverse square of the distance, meaning that the field intensity diminishes with increasing distance from the source. In the vicinity of the BH, the strong gravitational pull causes the charged particles to accelerate significantly, which clearly strengthens the magnetic field. Additionally, dynamo effects occurring near the BH due to rotational motion and high acceleration contribute to increasing the field. However, as the distance from the BH increases, these effects diminish, and the magnetic field lines become more dispersed. Therefore, this reduction in field strength is due to the decrease in field intensity with distance, the reduction of dynamo effects, and the scattering of field lines, which cause the radial and polar components to weaken with distance from the BH.

In these calculations, we observe that in the innermost region of the sub-Keplerian accretion flow, the intensity of the magnetic field lines is denser than in the Keplerian flows. Consequently, it becomes evident that the magnetic field strength in the sub-Keplerian flow surpasses that in the Keplerian flow, given the rotational velocities of the flow. However, it is clear that the magnetic field strength for both keplerian and sub-Keplerian flows remains below the Eddington magnetic field limit when using Newton's solution, and the magnetic field does not reach its maximum capacity in the distant region from the BH. Yet, due to the significant growth of the magnetic field components and the accumulation of magnetic flux from an outer radius of $220 r_{g}$ to a radius of $20 r_{g}$, it can be concluded that the process of strengthening and growing the magnetic field has progressed to the point where the disc, from its initial conditions, is evolving toward the MAD state. These evolutions indicate that the disc is in a region close to the MAD state, where the magnetic forces are strong enough to balance the gas forces. This sub-MAD state suggests that the disk has not yet reached maximum magnetic flux accumulation, but dynamical processes continue to strengthen the magnetic field. A more robust magnetic field corresponds to a higher density of magnetic energy in the surrounding medium. Consequently, following the equipartition between the kinetic energy of the falling matter and the magnetic energy during the accretion, as the density of magnetic energy cannot exceed that of the kinetic energy of flow, any surplus is dissipated as heat into the environment. Based on these findings, it can be asserted that the quantity of heat released in the sub-Keplerian flow significantly surpasses that in the Keplerian flow, as we intend to elucidate in subsequent articles.

Before our work, \citeauthor{1974Ap&SS..28...45B} investigated the behavior and growth of the magnetic field around a non-rotating BH in the Schwarzschild metric. In the absence of rotational velocity in the flow, they found the magnetic field magnitude to $1 G$ around the pole and $0.1 G$ around the equator. When considering the $u^\varphi$ component of the flow, the growth of the magnetic field for both Keplerian and sub-Keplerian flows is greater than what they reported. Thus, in the presence of the $u^\varphi$ component, the behavior of the magnetic field aligns more closely with simulation results.

The rotational velocity component plays a crucial role in creating and strengthening the magnetic field. Our results indicate that the presence of the $u^\varphi$ velocity component in Keplerian flows leads to a high rotation speed, which in turn generates stronger magnetic dynamos that significantly increase the $_r B$ and $_\theta B$ components. Therefore, the magnetic field grows more rapidly, which can be attributed to the presence of $u^\varphi$.
It is important to note that the $u^\varphi$ component plays a significant role not only in Keplerian flows but also in sub-Keplerian flows. While the rotation speed in sub-Keplerian flows is generally lower, it still contributes to the magnetodynamic process and affects the growth and distribution of the magnetic field. Our findings show that despite the lower $u^\varphi$ in sub-Keplerian flows, the resulting magnetic field strength can be significant, reaching about $30 G$ around the pole. This highlights the necessity of considering the $u^\varphi$ velocity component in both types of flows to fully understand the dynamics of the magnetic field. 
Although one might expect that Keplerian flows, with their higher rotation speeds, would generate stronger magnetic fields, our results show that the sub-Keplerian flow exhibits a larger magnetic field strength. (According to Figures \ref{fig:fig1} and \ref{fig:fig4}, the magnitude of the sub-Keplerian flow field around the pole is $30 G$, while the magnitude of the Keplerian flow field around the pole is $12 G$). This difference can be attributed to variations in physical parameters such as density and viscosity. Moreover, while the higher rotation speed in the Keplerian flows enhances the dynamo effect, it may also increase energy dissipation, which could decrease the strength of the net magnetic field.
In conclusion, our results, which take into account the rotational velocity component, emphasize the critical importance of including it in the analysis of magnetic field dynamics. Our findings demonstrate that the $u^\varphi$ component significantly affects the strength and distribution of the magnetic field in both Keplerian and sub-Keplerian flows, underscoring its essential role in accurately modeling and understanding magnetohydrodynamic processes around BHs.



\bibliographystyle{mnras}
\bibliography{ref}




\appendix
\section{Appendix information}
To calculate the constants $c_{2}r_{g}^2$ and $c_{3}r_{g}^2$ for Keplerian and sub-Keplerian flows, it is necessary to insert the velocity components obtained for each flows separately in relation~(\ref{eq:eq9}). It should also be noted that we rewrite $B^{r}$ and $B^{\theta}$ in relation~(\ref{eq:eq9}) at $t=0$
\begin{equation}
\begin{split}
    B^{r} \left(t=0\right)  =& \ B_{0} \cos \theta   \sqrt{\left(1-x_0\right)},  \\
    rB^{\theta } \left(t=0\right)  =& -B_{0} \sin  \theta   \sqrt{\left(1-x_0\right)}.
\end{split}
\label{eq:eqA1}      
\end{equation}

\begin{equation}
    \sqrt{-g}  u_{0}^{-1} B^{r} \left(u_{r} u^{r}+u_{0} u^{0}\right) = c_{2} r_{g}^{2},   \quad
    \sqrt{-g}  u_{r} B^{\theta } = c_{3} r_{g}^{2}.
    \label{eq:eqA2} 
\end{equation}
\\
Keplerian flow:
\begin{equation}
\begin{split}
    &u_{0} = 1,\quad u^{0} = \frac{1}{1-x},   \\
    &u^{r} = -\left(x-\frac{x \sin ^{2} \theta}{1-x} \right)  ^{1/2}, \quad 
    u_{r} =  \frac{1}{1-x}  \left(x-\frac{x \sin ^{2} \theta}{1-x} \right)^{1/2}.
\end{split}
\label{eq:eqA3}
\end{equation}
\begin{equation}
    \left(\frac{r_{g}}{x_0}\right) ^2  B_{0} \sin \theta  \cos \theta     \left( \frac{1}{1-x_{0}}\right)^{1/2}   \left[1- \left( x_{0} - \frac{x_{0}\sin ^{2}\theta }{1-x_{0}}    \right)   \right] = c_{2} r_{g}^2.
    \label{eq:eqA4} 
\end{equation}
\begin{equation}
    \left(\frac{r_{g}}{x_0}\right)  B_{0} \sin ^2\theta      \sqrt{1-x_{0}}    \left[ x_{0} - \frac{x_{0}\sin ^{2}\theta }{1-x_{0}}   \right]^{1/2} = c_{3} r_{g}^2.
    \label{eq:eqA5} 
\end{equation}
Also, $c_{1}r_{g}$ for Keplerian flow is obtained by solving equation~(\ref{eq:eq21}) at $t = 0$.
\begin{equation}
    ct + Bx + C x^{3/2} + D x^{-3/2} + E x^{2} = c_{1} r_{g}.
    \label{eq:eqA6} 
\end{equation}
\begin{equation}
    c_{1} r_{g}= Bx_{0} + C x_{0}^{3/2} + D x_{0}^{-3/2} + E x_{0}^{2}.
    \label{eq:eqA7} 
\end{equation}
\\
Sub-Keplerian flow:
\begin{equation}
\begin{split}
    &u_{0} = 1,\quad u^{0} = \frac{1}{1-x},   \\
    &u^{r} = -\left(\frac{x \left(1-x\right) \left(r_{g}\sin \theta\right)  ^{2}}{x^{2} + \left(1-x\right) \left(r_{g}\sin\theta\right)  ^{2} } \right)  ^{1/2}, \\
    &u_{r} =  \left(\frac{x \left(r_{g}\sin \theta\right)  ^{2}}{x^{2}\left(1-x\right)  + \left(1-x\right)^{2} \left(r_{g}\sin\theta\right)  ^{2} } \right)  ^{1/2}.
    \end{split} 
\label{eq:eqA8} 
\end{equation}
\begin{equation}
\begin{split}
    &\left(\frac{r_{g}}{x_0}\right) ^2  B_{0} \sin \theta  \cos \theta     \sqrt{1-x_{0}}   \\
    &\left[\frac{1}{1-x_{0}} - \frac{x_{0} \left(r_{g}\sin \theta \right) ^{2} }{x_{0}^{2}+\left(1-x_{0}\right) \left(r_{g}\sin \theta \right) ^{2}  }    \right] = c_{2} r_{g}^2.
    \end{split}
\label{eq:eqA9} 
\end{equation}
\begin{equation}
    \left(\frac{r_{g}}{x_0}\right)  B_{0} \sin^{2} \theta  \    \sqrt{1-x_{0}}   \left[ \frac{x_{0} \left(1-x_{0}\right)  \left(r_{g}\sin \theta \right) ^{2} }{x_{0}^{2}+\left(1-x_{0}\right) \left(r_{g}\sin \theta \right) ^{2}  }    \right]^{1/2} = c_{3} r_{g}^2.
    \label{eq:eqA10} 
\end{equation}
Also, $c_{1}r_{g}$ for sub-Keplerian flow is obtained by solving equation~(\ref{eq:eq25}) at $t = 0$.
\begin{equation}
    ct + \frac{2}{3} x^{-3/2} + 2 x^{-1/2} + \ln \left( \frac{1 - \sqrt{x}}{1 + \sqrt{x}} \right) = c_{1} r_{g}.
    \label{eq:eqA11} 
\end{equation}
\begin{equation}
    c_{1} r_{g}=\frac{2}{3} x_0^{-3/2} + 2 x_0^{-1/2} + \ln \left( \frac{1 - \sqrt{x_0}}{1 + \sqrt{x_0}} \right).
    \label{eq:eqA12} 
\end{equation}


\bsp	
\label{lastpage}
\end{document}